\documentclass[a4paper,fleqn,usenatbib]{mnras}
\usepackage[T1]{fontenc}
\usepackage{ae,aecompl}
\usepackage{graphicx}
\usepackage{amsmath}
\usepackage{amssymb}
\usepackage{natbib}
\usepackage{relsize}
\usepackage{enumitem}
\usepackage[utf8]{inputenc}
\usepackage{epstopdf}
\usepackage{subfigure}
\usepackage{pifont}
\newcommand{\cmark}{\ding{51}}%
\newcommand{\xmark}{\ding{55}}%

\title[Rotation and Li depletion in young clusters]{The Gaia-ESO Survey: Constraining evolutionary models and ages for young low mass stars with measurements of lithium depletion and rotation.}

\author[A. S. Binks et al.]{A. S. Binks$^{1,2}$\thanks{E-mail: abinks@space.mit.edu}, R. D. Jeffries$^{1}$, G. G. Sacco$^{3}$, R. J. Jackson$^{1}$, L. Cao$^{4}$, A. Bayo$^{5,6}$, M. Bergemann$^{7,8}$,
\newauthor
R. Bonito$^{9}$, G. Gilmore$^{10}$, A. Gonneau$^{10}$, F. Jimin\'ez-Esteban$^{11}$, L. Morbidelli$^{3}$, S. Randich$^{3}$, 
\newauthor
V. Roccatagliata$^{12}$, R. Smiljanic$^{13}$, S. Zaggia$^{14}$
\\
$^{1}$MIT Kavli Institute for Astrophysics and Space Research, Massachusetts Institute of
Technology, Cambridge, MA 02139, USA\\
$^{2}$Astrophysics Group, School of Chemical and Physical Sciences, Keele University, Keele, ST5 5BG, UK \\
$^{3}$ INAF -- Osservatorio Astrofisico di Arcetri, Largo E. Fermi, 5, 50125 Firenze, Italy\\
$^{4}$ 4055 McPherson Laboratory, 140 West 18th Avenue, Columbus, Ohio 43210-1173, USA\\
$^{5}$ Instituto de F\'isica y Astronom\'ia, Fac. de Ciencias, U de Valpara\'iso, Gran Breta\"na 1111, Playa Ancha, Chile\\
$^{6}$European Southern Observatory (ESO), Karl-Schwarzschild-Str. 2, 85748, Garching b. M\"unchen, Germany\\
$^{7}$ Galaxies $\&$ Cosmology Department, Max Planck Institute for Astronomy, Koenigstuhl 17, Heidelberg, D-69117, Germany\\
$^{8}$ Niels Bohr Institute, University of Copenhagen, Blegdamsvej 17, DK-2100 Copenhagen, Denmark\\
$^{9}$ INAF -- Osservatorio Astronomico di Palermo, Piazza del Parlamento, 1 90134, Palermo, Italy\\
$^{10}$ Institute of Astronomy, University of Cambridge, Madingley Road, Cambridge CB3 0HA, UK\\
$^{11}$ Nicolaus Copernicus Astronomical Center, Polish Academy of Sciences, ul. Bartycka 18 00-716 Warsaw, Poland\\
$^{12}$ Dipartimento di Fisica ``Enrico Fermi'', Universita’ di Pisa, Largo Pontecorvo 3, 56127 Pisa, Italy\\
$^{13}$ Centro de Astrobiología (INTA-CSIC), P.O. Box 78, 28691 Villanueva de la Ca\"nada, Madrid, Spain\\
$^{14}$ INAF -- Osservatorio Astronomico di Padova, Vicolo dell’Osservatorio 5, I-35122 Padova, Italy
}
\date{Accepted 2022 April 10. Received 2022 March 23; in original form 2021 November 24}
\pubyear{2022}

\begin{document}
\label{firstpage}
\pagerange{\pageref{firstpage}--\pageref{lastpage}}
\maketitle

\begin{abstract}
A growing disquiet has emerged in recent years that standard stellar models are at odds with observations of the colour-magnitude diagrams (CMDs) and lithium depletion patterns of pre main sequence (PMS) stars in clusters. In this work we select 1,246 high probability K/M-type constituent members of 5 young open clusters (5--125\,Myr) in the Gaia-ESO Survey to test a series of models that use standard input physics and others that incorporate surface magnetic fields or cool starspots. We find that: standard models provide systematically under-luminous isochrones for low-mass stars in the CMD and fail to predict Li-depletion of the right strength at the right colour; magnetic models provide better CMD fits with isochrones that are $\sim 1.5-2$ times older, and provide better matches to Li depletion patterns. We investigate how rotation periods, most of which are determined here for the first time from Transiting Exoplanet Survey Satellite data, correlate with CMD position and Li. Among the K-stars in the older clusters we find the brightest and least Li-depleted are the fastest rotators, demonstrating the classic ``Li-rotation connection'' for the first time at $\sim 35$ Myr in NGC~2547, and finding some evidence that it exists in the early M-stars of NGC~2264 at $<10\,$Myr. However, the wide dispersion in Li depletion observed in fully-convective M-dwarfs in the $\gamma$~Vel cluster at $\sim 20$ Myr appears not to be correlated with rotation and is challenging to explain without a very large ($>10$ Myr) age spread.
\end{abstract}

\begin{keywords}
stars: kinematics and dynamics --- stars: late-type --- stars: pre-main-sequence --- (Galaxy:) solar neighbourhood
\end{keywords}

\section{Introduction}\label{S_Introduction}

Lithium (Li) in the photospheres of pre main sequence (PMS) stars is an incisive probe of their interior physics. As they contract towards the zero age main sequence (ZAMS), their cores attain temperatures high enough ($\sim 3\times 10^{6}$\,K) to initiate  $^7$Li(p,$\alpha$)$^{4}$He reactions. At low masses ($<0.6 M_{\odot}$), a fully convective interior ensures rapid mixing and Li destruction throughout the star, prior to arrival on the ZAMS \citep[e.g.][]{1997a_Bildsten}. In higher mass stars ($0.6<M/M_{\odot}<1.1$), a radiative core should develop before Li destruction is complete and prevents further mixing of undepleted material to regions where it can be ``burned'' \citep[e.g.][]{1997a_Pinsonneault, 2002a_Piau}. In these cases there should be partial, mass dependent, depletion of photospheric Li. Many recent reviews provide a description of the physics of Li-depletion in PMS low-mass stars \citep{2014a_Jeffries,2016a_Lyubimkov,2021a_Randich,2021a_Tognelli}.

The focus of most previous work to test these ideas has been in the F-, G- and K-type stars of young open clusters \citep{2020a_Semenova}. These co-eval groups of stars do exhibit the expected age- and mass-dependent progress of photospheric Li expected from ``standard'' models \citep[][]{1993a_Soderblom, 1993a_Thorburn}. However, there is a scatter in Li abundances at a given effective temperature ($T_{\rm eff}$), reaching 2 orders of magnitude for K-stars at the ZAMS \citep{1983a_Duncan, 1993a_Soderblom, 1998a_Randich, 1998a_Jeffries, 2001a_Randich}, which suggests that other parameters are influencing Li depletion and that are not included in the ``standard'' models.

It has become clear that whatever the additional mechanisms are, they leave an imprint in the form of a strong correlation between photospheric Li abundance and rotation. Following the work of \cite{1987a_Butler}, the connection between Li and rotation has been firstly and firmly established amongst K-stars in the Pleiades, where fast rotators tend to be more luminous and less Li depleted \citep{1993a_Soderblom,2000a_King,2015a_Somers,2016a_Barrado,2018a_Bouvier}. Subsequent work has confirmed this trend appears universal from clusters as young as 5\,Myr (NGC~2264: \citealt{2016a_Bouvier}) and possibly even at 3\,Myr ($\sigma$\,Ori: Garcia Villota 2022, in prep.), and persists up to ages at least $\sim 150$\,Myr (M35: \citealt{2021a_Jeffries}). The same trend is also found in sparser, nearby moving groups (MGs; e.g., the $\beta$ Pictoris MG: \citealt{2016a_Messina}), and in recently discovered young filamentary associations (Pisces-Eridanus: \citealt{2020a_Arancibia-Silva}). \cite{2021a_Bouma} recently used Li and rotation period ($P_{\rm rot}$) measurements of NGC~2516 members to confirm the existence of the 500\,pc-long tidal tails discovered by \cite{2021a_Meingast}, finding that the Li-rotation connection is enduring, even in the extended structures. \cite{2021a_Llorente_de_Andres} recently compiled Li and $v\sin i$ measurements for FGK-type members of dozens of clusters and associations and computed a $v\sin i$ threshold, separating Li-rich and Li-poor stars, which decreases with age. 

The predominant ideas to explain the Li-rotation connection are either: (i) PMS Li depletion is inhibited in fast rotators either as a result of dynamo-generated magnetic fields or other rotation-dependent manifestations of magnetic activity such as starspots, reducing convective efficiency and interior temperatures \citep[e.g.][]{2014b_Jackson, 2014a_Somers, 2015a_Somers} or through a rotation-dependent change in the stability criterion against convection \citep{2021a_Constantino}; (ii) PMS Li depletion is enhanced in slow rotators as a result of additional mixing processes associated with angular momentum loss or convective overshooting that are either more effective in slower rotators or become inhibited in fast rotators \citep[in those stars with a radiative core, e.g.][]{2012a_Eggenberger, 2017a_Baraffe}.

Understanding PMS Li depletion in low-mass stars is also important as a means of verifying the accuracy of evolutionary models, and as a way of validating the ages deduced for young stars, clusters and star forming regions derived from them. The position of a PMS star in the Hertzsprung-Russell diagram (HRD) and its level of Li depletion provide partially independent constraints. Isochrones in the HRD and of Li depletion should be matched at the same age for a valid model. That this is not the case has been brought into focus by work that shows ages derived from Li depletion in PMS low-mass stars are usually much older (often by a factor of two) than ages derived from fitting isochrones to higher mass, or even the same stars, in the HRD \citep{2016a_Feiden, 2017a_Jeffries, 2021a_Binks, 2022a_Franciosini}.

In this paper, we study of PMS Li depletion in order to explore the extent to which various models predict their Li depletion patterns and colour-magnitude diagrams (CMDs). To do this, we exploit data from the {\it Gaia}-ESO Survey \citep[GES,][]{2012a_Gilmore, 2013a_Randich}, a large spectroscopic programme carried out at the European Southern Observatory's Very Large Telescope (VLT), which obtained many thousands of spectra for stars in open clusters, measuring equivalent widths for the Li\,{\sc i}~6708\AA\ diagnostic feature (EW(Li), calculated independently in this work using GES spectra) and radial velocities (RVs). In this respect, our work is similar to the GES-based studies of \cite{2017a_Jeffries} and \cite{2022a_Franciosini}, who both found that models incorporating starspots provided a better match to the CMDs and Li depletion patterns in young clusters. Our work provides an independent study to test evolutionary models with open clusters, using similar GES data, but incorporating a consideration of how stellar rotation affects the observed CMD and Li-depletion patterns. Here we focus on lower mass PMS stars ($\leq 0.6 M_{\odot}$) and, by combining GES spectroscopy, Gaia photometry and new rotation periods obtained from analysis of data from the Transiting Exoplanet Survey Satellite \citep[TESS,][]{2015a_Ricker} and other published sources, for the first time we explore whether the connection between rotation and Li depletion is present among late-K and M-type PMS stars prior to their total Li depletion.

In $\S$\ref{S_Sample_Selection}~we describe the open clusters chosen for analysis, the selection of high-probability cluster members and how we estimate their Li content from GES data and obtain their rotation periods, either from published sources or newly determined from TESS archival data. In $\S$\ref{S_Analysis}~we test to what extent evolutionary models are capable of simultaneously fitting CMDs and Li depletion patterns for the nominated clusters and in $\S$\ref{S_Prot_Analysis}~we investigate the trends of stellar rotation with CMD position ($\S$\ref{S_Prot_CMD}) and Li depletion ($\S$\ref{S_Prot_Li}). The implications of our results are provided in $\S$\ref{S_Discussion}~and the outcomes of this work are summarised in $\S$\ref{S_Summary}.

\section{Sample selection, spectroscopy and rotation periods}\label{S_Sample_Selection}

\subsection{Selecting clusters}\label{S_Cluster_Selection}

Our study focuses on five clusters (NGC 2264, $\lambda$~Ori, $\gamma$~Vel, NGC~2547 and NGC~2516)  observed during the GES campaign. Their ages of 3--125\, Myr (adopted from column 2 in \citealt{2022a_Jackson}), distance moduli (see $\S$\ref{S_Target_Selection}) and reddening values (adopted from literature sources) are listed in Table~\ref{T_Clusters}. Establishing whether these ages are reliable is one of the aims of this work since these clusters may be important benchmarks to interpret the timescales upon which gas is depleted from primordial disks and rocky planets are formed.

These clusters were selected because: a) they represent important epochs of stellar evolution from the (mass-dependent) early PMS to the ZAMS, covering the main PMS Li depletion phase; b) the number of cluster members in GES is reasonably large ($\sim10^{2}-10^{3}$ in each); c) they are nearby ($d = 350-750\,$pc) enabling reliable $P_{\rm rot}$ measurements for their low-mass members (either from our own TESS lightcurve analysis -- see $\S$\ref{S_TESS}, or from literature sources). 

Finally, these clusters have homogeneously determined metallicities that are within $\sim 0.07$ dex of the solar value (and each other), minimising any composition-dependent effects on their HRD or Li depletion patterns \citep[see Table~\ref{T_Clusters}, Appendix~\ref{S_Appendix1} and][]{2014a_Spina, 2017a_Spina, 2017a_Magrini, 2020a_Randich}.

\subsection{Target selection}\label{S_Target_Selection}

{\centering
\begin{table*}
\caption{Properties of the clusters used in this paper. Column 2 gives a nominal age for each cluster and column 3 gives the spectroscopic resolving power of the Giraffe HR15N set-up -- both taken from table 1 in \protect\cite{2020a_Jackson}. Column 4 is the distance modulus, where the error bars correspond to the statistical and systematic uncertainty, respectively (determined in this paper -- see $\S$\ref{S_Target_Selection}). Column 5 gives the reddening value adopted from the literature, where the references are (respectively): \protect\cite{2012a_Turner}; \protect\citet[for its eponymous member --- an uncertainty of $0.05$ is adopted]{1994a_Diplas}; \protect\cite{2009a_Jeffries}; \protect\cite{2006a_Naylor}; \protect\cite{2002a_Terndrup}. Column 6 is the median iron abundance (see Appendix~\ref{S_Appendix1}). Columns 7 and 8 give the number of cluster members used in this work and the number which have a measured rotation period ($P_{\rm rot}$), respectively, where the values in parentheses for the latter show how many are first-time $P_{\rm rot}$ measurements using TESS archival data (see \S\ref{S_TESS}). Column 9 is an estimate of the likely number of contaminant non-members within $N_{\rm mem}$ (see $\S$\ref{S_Target_Selection}).}
\begin{tabular}{lrrrrrrrr}
\hline
\hline
Cluster       &    Age & $R$                       & $M-m$                       & $E(B-V)$          & [Fe/H]           & $N_{\rm mem}$ & $N_{\rm P_{\rm rot}}$ & $N_{\rm con}$ \\
              &  (Myr) & ($\lambda/\Delta\lambda$) & (mag)                       & (mag)             &                  &               &                       &               \\
\hline
NGC~2264      & $3$    & $14968$                   & $9.301 \pm 0.010 \pm 0.047$ & $0.075 \pm 0.060$ & $-0.04 \pm 0.01$ & $404$         & $229 (80)$                 & $2.84$ \\
$\lambda$~Ori & $6$    & $17281$                   & $7.991 \pm 0.018 \pm 0.026$ & $0.120 \pm 0.050$ & $-0.06 \pm 0.02$ & $166$         & $68 (68)$                  & $0.40$ \\
$\gamma$~Vel  & $5-10$ & $14301$                   & $7.740 \pm 0.034 \pm 0.023$ & $0.038 \pm 0.016$ & $+0.01 \pm 0.02$ & $188$         & $47 (47)$                  & $0.93$ \\
NGC~2547      & $35$   & $13862$                   & $7.947 \pm 0.009 \pm 0.025$ & $0.120 \pm 0.020$ & $-0.01 \pm 0.01$ & $123$         & $73 (11)$                  & $0.08$ \\
NGC~2516      & $125$  & $13440$                   & $8.074 \pm 0.006 \pm 0.027$ & $0.120 \pm 0.020$ & $-0.02 \pm 0.01$ & $365$         & $299 (49)$                 & $0.56$ \\
\hline
\end{tabular}
\label{T_Clusters}
\end{table*}}

The targets within the clusters were observed as part of GES. The initial target selection for low-mass stars was unbiased with respect to previous claims of membership, Li abundance or indications of magnetic activity and accretion. The main selection was photometric, encompassing stars from a range around a nominal cluster locus in CMDs, which was broad enough to encompass the likely positions of single stars and any multiple systems \citep[e.g.][]{2018a_Randich, 2022a_Bragaglia}. Almost all spectra were obtained with the Giraffe multi-object spectrograph using the HR15N setup ($\lambda\lambda 6444-6816\,$\AA) at a resolving power of $R\sim 15\,000$, with a few UVES echelle spectra ($R=47\,000$) obtained for brighter members of each cluster \citep[see][]{2014a_Sacco}. The median resolving power using the HR15N setup for each cluster (which varied over the course of the survey and was measured from calibration lamps) is provided in column 3 of Table~\ref{T_Clusters}.

With the exception of Li measurements (see $\S$\ref{S_LiEW}) all spectroscopic measurements and stellar parameters were taken from the sixth GES internal data release (GESiDR6, access available for consortium members only\footnote{\url{http://gaia-eso.eu/}}, which are based on the GES DR4 spectra, made publicly available in the ESO archive\footnote{\url{http://archive.eso.org/cms/eso-data.html}}). Valid targets for this paper were required to have 3D kinematic membership probabilities $P_{\rm 3D} \geq 0.95$ (calculated in \citealt{2022a_Jackson}; $P_{\rm 3D} \geq 0.97$ was used for NGC~2264, where levels of background contamination were higher, see Table~\ref{T_Clusters}) and $\log g > 3.0$ (to avoid contamination from giant stars). The young stars around $\gamma$~Vel consist of at least two populations with similar ages and Li-depletion patterns but slightly differing kinematics, separated by a distance $\sim 40$\,pc \citep[populations A and B,][]{2014a_Jeffries,2018a_Franciosini,2020a_Armstrong}. In this work the $\gamma$~Vel sample consists of targets with $P_{\rm 3D} > 0.95$ of being associated with (the foreground) population A, such that contamination from members of population B is very low. Table~\ref{T_Data}~provides $P_{\rm 3D}$ and EW(Li) values for objects that pass all the selection criteria described in this section.

The membership probabilities given by \cite{2022a_Jackson} are primarily kinematic, based on astrometric data from the Third (Early) Gaia Data Release (GEDR3, \citealt{2020a_Gaia_Collaboration})~and RVs from GES (i.e. Li abundance or position in the CMD play no role in membership selection). GEDR3 provides significant astrometric precision improvements over the Second Gaia Data Release (GDR2) \citep{2018a_Gaia_Collaboration}, however we used $G$ band photometry from GDR2 because the photometric passbands used to predict outputs in the evolutionary models we discuss later are those of GDR2, which have subtle differences from those of GEDR3 (see $\S$\ref{S_Analysis_CMD}). We use the GEDR3 parallaxes to calculate distance moduli for each cluster following the method described in \cite{2021a_Binks}, where we adopt a systematic parallax uncertainty $=0.03\,$mas, based on figure 2a in \cite{2021a_Lindegren}. The median difference between the GDR2 and GEDR3 parallax measurements $=-0.025$\,mas, with a scaled median absolute deviation $=0.062$\,mas, and our measurements of distance modulus ($M-m$) for each cluster are within 0.02\,mag of those provided in \cite{2022a_Jackson}, where the systematic errors calculated in both works are almost identical. Distance moduli are provided in Table~\ref{T_Clusters} with their statistical and estimated systematic error bars.

In $\S$\ref{S_Analysis}~we construct optical/infrared CMDs, where we use GDR2 $G$ magnitudes exclusively as the optical component. For each GDR2 source identifier, we perform a cross-match with targets in the 2MASS~photometric catalogue \citep{2003a_Cutri} and in the sixth data release of the Vista Hemisphere Survey (herein, VHS, \citealt{2021a_McMahon}) to within 2.0'' of their position in GDR2 (at the J2000 epoch). We obtain $K_{\rm s}$ band photometry for all targets in 2MASS, and for all members of $\gamma$~Vel, NGC2547 and NGC~2516 in VHS (the VHS sky coverage is limited to $\delta < 0^{\circ}$), however the bright targets are saturated at $K_{\rm s}$ in VHS, therefore we decided to use only 2MASS data, which ensures the $K_{\rm s}$ magnitudes are from a homogeneous source. Two objects have a 2MASS~$K_{\rm s}$ quality flag ($Q_{\rm K_{\rm s}}$) = `U', corresponding to badly calibrated photometry, and were discarded.

For the purposes of this paper we focused on stars with spectral types between K0 and M5. Objects are further separated into spectral-type bins: K0$-$K5, K5$-$M0, M0$-$M2 and M2$-$M5, where the bin boundaries for the four youngest clusters (NGC~2264, $\lambda$ Ori, $\gamma$ Vel and NGC~2547) are obtained by linear interpolation of $(G-K_{\rm s})_0$ versus spectral-type from table 6 of \citet{2013a_Pecaut}, where $G-K_{\rm s}$ is converted from $V-I_{\rm c}$ using the relations provided by J. M. Carrasco in table 5.8 of the GDR2 online documentation\footnote{\url{https://gea.esac.esa.int/archive/documentation/GDR2/Data_processing/chap_cu5pho/sec_cu5pho_calibr/ssec_cu5pho_PhotTransf.html}}. For the older NGC~2516 cluster, a direct conversion between $G-K_{\rm s}$ and spectral-type is made using an online main-sequence conversion table provided by E. Mamajek\footnote{\url{https://www.pas.rochester.edu/~emamajek/EEM_dwarf_UBVIJHK_colors_Teff.txt}}. The $G-K_{\rm s}$ colours for each cluster member are dereddened using the cluster value of $E(B-V)$ in Table~\ref{T_Clusters} and the $G$ and $K_{\rm s}$ extinction relations in \cite{2019a_Wang}. The uncorrected $G$ and $K_{\rm s}$ photometry and de-reddened spectral-types for each object are provided in Table~\ref{T_Data}. Uncertainties in $E(B-V)$ lead to spectral-types with a precision of about half a sub-class for M-stars (e.g., M2V versus M2.5V) and one sub-class for K-stars.

These selection criteria provide a catalogue of 1,\,246 objects, where the number in each cluster, $N_{\rm mem}$, is listed in Table~\ref{T_Clusters}. The estimated number of contaminants among these high probability cluster members is calculated as $N_{\rm c} = (1 - \overline{P_{\rm 3D}})\times{N_{\rm mem}}$, where $\overline{P_{\rm 3D}}$ is the mean membership probability of the selected members. NGC~2264 has the lowest value of $\overline{P_{\rm 3D}}= 0.993$ and the largest value of $N_{\rm c}= 2.8$, probably because it has the broadest intrinsic velocity dispersion of these clusters \citep{2020a_Jackson}.

\subsection{Identifying strongly accreting targets}\label{S_Accretors}

When comparing rotation periods and EW(Li) later in this work ($\S$\ref{S_Prot_Analysis}), we discard any strongly accreting stars (though they will be retained for all other purposes), as
 these may have strongly veiled continua \citep{2003a_Stempels}, leading to erroneously small EW(Li) values. A common method to identify strong accretors is to measure the velocity profile of the H$\alpha$ feature \citep{2003a_White}. Indeed, \cite[][herein, B20]{2020a_Bonito} identified strong accretors in NGC~2264 using detailed line profile information from the same GES spectroscopic data used in this work.

We used a simpler way to identify strong accretors since most targets in our sample lack detailed H$\alpha$ profile measurements in the GESiDR6 data, and the SNR values of our faintest targets are low, leading to large error bars, or potentially spurious velocities. Instead, we define strong accretors, using a modified version of eqn.~1 in \cite{2003a_Barrado_y_Navascues}, as those with chromospheric H$\alpha$ EW, taken from the GESiDR6 recommended working group parameters, EW(H$\alpha_{\rm CHR}) > (5.6(G-K_{\rm s})_0 - 5.5)$\AA. Fifty per cent of the targets are missing a EW(H$\alpha_{\rm CHR}$) measurement and for the purposes of this work we assume they are not accreting stars.

All 404 NGC~2264 members in our catalogue are present in table 1 of B20, where B20 label 16 as ``good'' (clear signs of accretion), 270 as ``intermediate'' (spectroscopic quantities satisfy a given threshold) and 118 as ``bad'' (no signs of accretion). We find, for the ``good'' sample, that in all cases where a EW(H$\alpha_{\rm CHR}$) value is available (14 out of 16) that EW(H$\alpha_{\rm CHR}) >> 5.6(G-K_{\rm s})_0 - 5.5$, in complete agreement. For the ``intermediate'' sample, 210 stars with EW(H$\alpha_{\rm CHR})$ are available, where we identify 67 and 143 stars as accreting and non-accreting, respectively. Finally, for the ``bad'' sample we have 60 stars with a EW(H$\alpha_{\rm CHR})$ value, of which all but one are classed as non-accretors. Whilst we are confident our method does not wrongly categorise weak accretors, it is always possible that further strong accretors remain undetected because accretion is a highly variable process and we have missing data. We only find strong accretors in the two youngest clusters, with 29 and 10 (of the sample with rotation periods, see \S\ref{S_Prot_Overview}) in NGC~2264 and $\lambda$~Ori, respectively. This is to be expected since accretion has almost terminated by 10\,Myr in low-mass stars \citep{2010a_Fedele,2020a_Manara,2020a_Somigliana}.

{\centering
\begin{table*}
\begin{center}
\begin{tabular}{llllllllll}
\hline
\hline
Index & Name                     & Cluster  & $P_{\rm 3D}$ & $G$                  & $K_{\rm s}$        & SpT  & EW(Li)       & EW(H$\alpha_{\rm CHR}$) & $v\sin i$ \\
 & (Gaia DR2-)              &          &              & (mag)                & (mag)              &      & (m\AA)       & (\AA)                   & km\,s$^{-1}$ \\
\hline

1 &  3326610745242087040 & NGC~2264 & $0.9952$ & $16.5676$ & $12.384$ & M4.4 & $512 \pm 33$ &  $7.523 \pm 0.428$ & $15.0 \pm 0.7$ \\
2 &  3326895170860918912 & NGC~2264 & $0.9776$ & $14.9707$ & $11.890$ & K5.5 & $461 \pm 34$ &  $2.394 \pm 0.290$ & $15.0 \pm 0.9$ \\
3 &  3326908090122747904 & NGC~2264 & $0.9776$ & $15.8018$ & $12.919$ & K4.3 & $ 27 \pm 28$ &                    & $20.1 \pm 1.7$ \\
4 &  3326609985036956160 & NGC~2264 & $0.9911$ & $16.6417$ & $12.934$ & M2.7 & $571 \pm 25$ &  $4.356 \pm 0.330$ & $15.0 \pm 0.7$ \\
5 &  3326900840217848960 & NGC~2264 & $0.9981$ & $15.4001$ & $12.093$ & M0.0 & $553 \pm 27$ & $13.514 \pm 0.419$ &  $7.0 \pm 0.8$ \\
6 &  3326605797439777280 & NGC~2264 & $0.9963$ & $16.7418$ & $13.048$ & M2.6 & $544 \pm 28$ &                    & $18.0 \pm 0.7$ \\
7 &  3326705608184540928 & NGC~2264 & $0.9973$ & $17.5648$ & $13.728$ & M3.3 & $662 \pm 32$ &                    &                \\
8 &  3326705608184540160 & NGC~2264 & $0.9937$ & $17.2870$ & $13.417$ & M3.5 & $549 \pm 30$ &  $3.390 \pm 0.385$ & $24.0 \pm 0.7$ \\
9 &  3326907059330544640 & NGC~2264 & $0.9806$ & $14.6502$ & $11.837$ & K3.9 & $  8 \pm 37$ &                    & $28.7 \pm 1.8$ \\
10 & 3326908639878617216 & NGC~2264 & $0.9928$ & $15.1725$ & $11.901$ & K9.7 & $527 \pm 26$ &  $2.830 \pm 0.257$ & $41.1 \pm 1.6$ \\
\hline
\end{tabular}
\caption{Collated data for the 1,\,246 stars in the sample. Column 1 provides a short reference for the corresponding targets in Tables~\ref{T_TESS}~and~\ref{T_Periods}, and columns 2 and 3 give the Gaia DR2 source identifiers and host clusters, respectively. Columns $4-10$ are as follows: membership probabilities (from \citealt{2022a_Jackson}), $G$ and $K_{\rm s}$ photometry, de-reddened photometric spectral-types, Li equivalent width measurements (EW(Li)) derived from our independent analysis of GESiDR4 spectra (see $\S$\ref{S_LiEW}), chromospheric H$\alpha$ equivalent widths (EW(H$\alpha_{\rm CHR}$)) and projected rotational velocities ($v\sin i$), where the latter two measurements are the recommended astrophysical parameters from GESiDR6. Typical $G$ and $K_{\rm s}$ uncertainties are 0.001 and 0.03 mag, respectively. Only the first 10 objects are displayed. The entire table is available in electronic format.}
\label{T_Data}
\end{center}
\end{table*}}

\subsection{Li equivalent widths}\label{S_LiEW}

There are significant difficulties in establishing the continuum level around the Li~{\sc i} line in cool targets, particularly with $T_{\rm eff}<4000$\,K, because molecular absorption features are present that can masquerade as Li absorption.
The EW(Li) measurements reported for M-dwarfs in GESiDR6 are ``pseudo equivalent widths" that include a significant, rotation-dependent molecular contribution, especially in sources with little lithium, and which are converted to abundances using a curve of growth that attempts to account for this contamination and rotation using synthetic spectra. Since in this paper we need to isolate the contribution from lithium to meet the key aim of looking for correlations with rotation and CMD position, we have remeasured EW(Li) for all the targets in this work, using the specialised analysis technique described in \cite{2021a_Binks}.

Briefly, EW(Li) is estimated by direct flux integration of the heliocentrically corrected spectrum and then subtracting the continuum flux using a fully Li-depleted template spectrum constructed from field stars of similar temperature. This should account for any molecular contribution to absorption. The only modification adopted here is that we used a top-hat function in the extraction, with a width tuned to $\pm v \sin i$ estimated for the star plus the instrumental resolution width, instead of a Gaussian profile. This is because there is some additional uncertainty about the line profile shapes for the youngest stars in our sample with $\log g < 4.0$, and their continua are less well matched to the empirically-derived continuum from the background stars, which are mostly main sequence dwarfs. The top-hat extraction is less precise (i.e. the error bars are slightly larger), but should provide better absolute accuracy.

Simulations performed by \cite{2016a_Bouvier} find that amongst Li-rich stars with $v\sin i<120\,{\rm km\,s}^{-1}$ any systematic offset in EW(Li) measurement caused by fast rotation is no more than a few 10s of m\AA~(see their figure A.1). Only 17 targets in our sample have $v\sin i>120\,{\rm km\,s}^{-1}$ as measured in GESiDR6, only 2 of which rotate faster than $200\,{\rm km\,s}^{-1}$. Since these offsets are small compared to the measurement uncertainties (particularly for the fast-rotating M-stars) and our technique for measuring EW(Li) should take account of rotational broadening and blending in any case, the systematic uncertainty should be negligible. All EW(Li)s estimated for the targets are presented in Table~\ref{T_Data}.

\subsection{Rotation periods}\label{S_Prot_Overview}
In $\S$\ref{S_Prot_Analysis}~we examine how rotation affects luminosity and Li-depletion by constructing a sample of several hundreds of rotation periods ($P_{\rm rot}$). As well as collecting data from previous works, we have developed a technique to estimate $P_{\rm rot}$ for stars in our target sample using the 30-minute cadence time-series observations from TESS \citep{2015a_Ricker}. Briefly this involves: (a) extracting data from the {\sc tesscut} tool hosted at MAST (\citealt{2019a_Brasseur}\footnote{\url{https://mast.stsci.edu/tesscut/}}); (b) preprocessing the lightcurve data; (c) implementing a periodogram analysis to measure $P_{\rm rot}$; (d) developing reliability and quality criteria based on various features of the lightcurve and contamination from background sources and (e) a procedure to calculate $P_{\rm rot}$ when there are multiple seasons of reliable TESS data and/or previous $P_{\rm rot}$ measurements from literature sources.

The size of a TESS pixel is 21", which can lead to a significant flux contribution from contaminating sources. This presents additional difficulties for some targets, especially in NGC~2264 because the cluster is relatively dense. By cross-matching neighbouring sources within a radius of 5 TESS pixels in the Gaia DR2 catalog, that may contribute to the observed flux in the TESS light curve, we are able to identify this contamination. All the methods used to calculate TESS-based $P_{\rm rot}$ values, efforts to identify flux contamination and the source of all the period data used in this work are outlined in Appendix~\ref{S_Rotation_Periods}.

The comparison between TESS periods and those from the literature in Figure~\ref{F_Period_Comparison}~looks reasonable, with 73 per cent of these objects having consistent measurements (as defined in Appendix~\ref{S_TESS_Comparison}). We note that many of the literature sources are from ground based photometry, and may be more prone to 1-day aliasing issues. Approximately half of the discrepant periods appear to be half, or double values of each other, and in these cases we select the measurement with the largest $P_{\rm rot}$, since the shorter period could be a manifestation of spots on both hemispheres of a star. Visual assessment of the light curves show the remaining discrepant periods have noisy light curves. Recently, \cite{2021a_Bouma} performed a similar TESS analysis for a large sample of NGC~2516 candidate members, where 108 objects in their representative high-quality sample are present in our analysis. There is almost perfect agreement for $P_{\rm rot}<6\,$days and reasonable agreement at longer periods, with some evidence of a systematic offset at longer periods of about 1--2 days (our periods are shorter), but most of these targets have very low-amplitude, noisy light curves.

By counting the number of stars with discrepant $P_{\rm rot}$ values, either amongst multiple TESS lightcurves, or where there are $\geq2$ values at odds with each other from literature sources (see $\S$\ref{S_Prot_Reliable}), we estimate that $\leq16$ per cent of the $P_{\rm rot}$ values used in this work may be unreliable. This fraction is an upper limit since the final chosen $P_{\rm rot}$ is likely to be correct for a substantial fraction of these stars.

Table~\ref{T_Clusters}~provides the number of targets in each cluster with a $P_{\rm rot}$ value (716 in total), almost three-quarters of which are in either NGC~2264 or NGC~2516. The values in parentheses give the number of targets that have $P_{\rm rot}$ value measured for the first time in this work (255 in total), which include members of $\lambda$~Ori and $\gamma$~Vel, representing the first $P_{\rm rot}$ measurements for {\it any} targets in these two clusters. Results from our periodogram analysis and the final $P_{\rm rot}$ values used in this work are presented in Tables~\ref{T_TESS}~and~\ref{T_Periods}, respectively.

\section{Can evolutionary models simultaneously describe the CMD and lithium depletion?}\label{S_Analysis}

The collected database of photometry, EW(Li) and (where available) rotation periods are used to test PMS evolutionary models, and where there are discrepancies, to see whether these discrepancies are related to rotation rate.

\subsection{Calibrating the evolutionary models}\label{S_Analysis_CMD}
\subsubsection{Constructing the isochronal fits in the CMD}
A first step is to compare the positions of all the cluster members with PMS models in both the $(G-K_{\rm s})_0/M_G$ intrinsic colour/absolute magnitude diagram (CMD), which serves as an observational proxy of the HRD, and the EW(Li) versus $(G-K_{\rm s})_0$ plane, which is an observational proxy for Li abundance versus $T_{\rm eff}$. Comparison is made in these observational diagrams because it makes the role of observational uncertainties and systematic errors clear. The data in Table~\ref{T_Clusters} are used to correct the observational data in each cluster for (an assumed uniform) reddening, extinction (see $\S$\ref{S_Target_Selection}) and distance modulus. The distance moduli are very precise but there is a potential additional systematic error (the second error bar in Table~\ref{T_Clusters}) associated with correlated uncertainties in the GEDR3 parallaxes on the angular scales of these clusters \citep[see][]{2021a_Lindegren}.

Beyond the photometric uncertainties, there are additional factors that could produce dispersion in CMD position. We have assumed a common distance for cluster members. In reality they are likely to be spread over a few pc, leading to a few hundredths of a magnitude errors in absolute magnitude. Similar dispersions are likely to be associated with modulation by starspots. There may also be some scatter introduced by neglecting differential extinction (particularly for the youngest clusters). Extinction amongst solar-type and high-mass members of NGC~2264 can vary up to a few tenths of a magnitude \citep{1978a_Young}, however the equivalent data for K/M-stars are lacking. It is also possible that a few disk-bearing M-stars are slightly under-luminous due to light scattering, particularly if these disks have an edge-on morphology. \cite{2012a_Bayo}~report that $\sim 60$ per cent of M-stars in $\lambda$ Ori have warm dusty disks, of which perhaps $\sim 10$ per cent are likely to be edge-on.

Three ``standard'' low-mass stellar evolutionary models are considered: \cite{2008a_Dotter}, \cite{2015a_Baraffe} and the spot-free version of the \cite{2020a_Somers} models. By ``standard'' we mean those models that do not include effects such as rotation or magnetic activity. The ``magnetic models'' are models that attempt to incorporate some aspects of the dynamo-generated magnetism which is known to be present in these young, fast-rotating, magnetically active stars. We consider two such evolutionary codes: (a) The SPOTS models of \citet[herein, S20]{2020a_Somers}, where a fraction of the photospheric flux (denoted $\beta$) is blocked by dark, magnetic starspots, with a spot/photoshere temperature ratio (denoted $\tau$) of 0.9\footnote{The S20 models with $\tau=0.9$ were provided by Lyra Cao (private communication).}. (b) The ``magnetic Dartmouth models'' described in \cite{2014a_Feiden} and \citet[herein, F16]{2016a_Feiden}, which implement magnetic inhibition of convection constrained by a boundary condition of an average $2.5\,$kG (roughly equipartition) field at the stellar surface. The F16 models can be considered an extension of the \cite{2008a_Dotter} standard models.

The $(G-K_{\rm s})_{0}$ and $M_{G}$ values from each model were uniformly calculated from cubic relationships between $\log T_{\rm eff}$ and $G-K_{\rm s}$, and between $\log T_{\rm eff}$ and the $G$ band bolometric corrections, at ages between 1.0 and 250.0\,Myr (in steps of 0.1\,Myr) from the \cite{2015a_Baraffe} models. This method was chosen to avoid problems in converting $T_{\rm eff}$ to colour and luminosity to $M_{\rm G}$ from (pre-)MS interpolation tables, and the \cite{2015a_Baraffe} models were selected specifically because they consistently couple realistic atmospheres and interior structure. The S20 (spotted) models require more careful treatment because they have two-temperature components (a hotter photospheric surface and a cooler spot region). Composite bolometric corrections in the $G$ and $K_{\rm s}$ bands were calculated using equation 6 from \cite{2014b_Jackson}.

The data and models are compared in the CMD using the fitting method described in detail by \cite{2021a_Binks}. Each cluster contains a fraction of multiple stellar systems, which if included in the fit would bias the isochronal fits to younger ages, since they appear brighter than their single counterparts \citep{2014a_Kamai}. We define a subset of likely-single cluster members by splitting the cluster population into two halves that lie either side of a second-order polynomial fit to the entire population in the CMD. The idea here is that the fainter half is representative of the single star population, whilst the brighter half is likely to consist mainly of unresolved binary systems. This separation into faint and bright samples will be used in the subsequent analysis in $\S$\ref{S_Prot_Analysis}. The choice of a 50 per cent ``binary'' fraction is supported by recent work that suggests the multiplicity fraction of K/M-type stars is larger than previously reported ($=0.46\pm0.06$, \citealt{2022a_Susemiehl}), and is similar to the value found in samples of higher-mass stars. The best-fit age corresponds to the isochrone that minimises $\chi^2 = \Delta^{2}/(N-1)$, where $\Delta^{2}$ is the sum of squared residuals in $M_{G}$ and $N$ is the number of targets used in the fit and this can be used as a figure of merit to compare fits with different models. 

To calculate a statistical uncertainty in age, we normalise $\Delta^2$ so that $\Delta^2_{\rm min}/(N-1) =1$ and then find the ages for which the normalised $\Delta^2 = N$ \citep[see][]{2021a_Binks}.
Statistical errors are $\lesssim 1$ Myr for the three youngest clusters and generally $1-2$\,Myr for the two oldest clusters, where more stars have reached the ZAMS causing a reduction in age sensitivity. The left-hand panels of Figures~\ref{F_SSM}~and~\ref{F_NSSM}~show best-fitting isochrones for each of the considered clusters in the CMD, where the $\chi^{2}$ versus age plots are shown in the insets. There are bigger systematic age uncertainties caused by the degeneracies between age, distance modulus and reddening, in the sense that a larger distance and a smaller reddening lead to younger estimated ages. Since the uncertainties in distance modulus and reddening are independent, their effects are added in quadrature and this is the dominant source of age uncertainty in most cases. The best-fit ages from all models used in this analysis, with statistical and systematic errors, and $\chi^{2}_{\rm min}$ values are reported in Table~\ref{T_CMD_ages}. 

The binary fraction among low-mass stars is unlikely to be larger than $\sim 50$ per cent \citep{2010a_Raghavan}~and could be smaller \citep[$\sim 25-30$ per cent, e.g.,][]{2013a_Duchene}. We tested the impact of this by offsetting the dividing line in the CMD until just 25 per cent of stars were discarded from the fits as binaries. The resulting isochronal fits (see column 4 in Table~\ref{T_CMD_ages}) are systematically younger, but by less than the systematic age error bar in most cases and so the exact choice of binary star fraction is not a dominant source of age uncertainty and would most likely affect the ages of all these clusters in a similar way.

\begin{table}
\begin{center}
\caption{Ages derived from fitting model CMDs from ``standard'' ($\S$\ref{S_SSM}: B15 = \citealt{2015a_Baraffe}; D08 = \citealt{2008a_Dotter}; S20 = the unspotted version of \citealt{2020a_Somers}) and ``magnetic'' models ($\S$\ref{S_NSSM}: F16 = \citealt{2016a_Feiden}; S20, with flux-blocking fractions, $\beta=0.1, 0.2$ and $0.3$) and a spot/photosphere temperature ratio, $\tau=0.9$, where the first error bar is statistical uncertainty from the $\chi^{2}$ fit and the second (larger) error bar is the age uncertainty due to the uncertainties in reddening and distance modulus for each cluster. Column 4 provides the best fit age where only 25 per cent of the sample are considered as binaries, column 5 shows the number of cluster members used in the fitting process ($N$) and column 6 gives $\chi^{2}_{\rm min}$ ($=\Delta^{2}_{\rm min}/(N-1)$, see $\S$\ref{S_Analysis_CMD}).}
\begin{tabular}{p{0.7cm}p{1.6cm}p{2.0cm}p{0.5cm}p{0.35cm}p{0.8cm}}
\hline
\hline
Cluster & Model & Age & Age$_{25}$ & $N$ & $\chi^{2}_{\rm min}$ \\
\hline

NGC           & B15              & $3.8\pm0.1\pm0.9$            & $3.0$   & $220$ & $0.2465$ \\
2264          & D08              & $4.6\pm0.2\pm0.9$            & $3.6$   & $190$ & $0.2245$ \\
              & S20, $\beta=0.0$ & $4.4_{-0.1}^{+0.2}\pm0.8$    & $3.6$   & $219$ & $0.2324$ \\
              & S20, $\beta=0.1$ & $5.2_{-0.1}^{+0.2}\pm1.0$    & $5.1$   & $218$ & $0.2042$ \\
              & S20, $\beta=0.2$ & $6.2\pm0.3\pm1.3$            & $4.9$   & $217$ & $0.2144$ \\
              & S20, $\beta=0.3$ & $8.5_{-0.7}^{+0.1}\pm1.9$    & $6.3$   & $219$ & $0.2053$ \\
              & F16              & $7.2_{-0.4}^{+0.1}\pm1.8$    & $5.5$   & $159$ & $0.2248$ \\
\hline
$\lambda$~Ori & B15              & $4.5\pm0.2\pm0.8$            & $3.9$   & $95$  & $0.1936$ \\
              & D08              & $5.6_{-0.1}^{+0.2}\pm1.1$    & $4.9$   & $87$  & $0.1538$ \\
              & S20, $\beta=0.0$ & $5.5\pm0.2\pm0.9$            & $4.9$   & $88$  & $0.1609$ \\
              & S20, $\beta=0.1$ & $6.4\pm0.2\pm1.9$            & $5.6$   & $91$  & $0.1528$ \\
              & S20, $\beta=0.2$ & $7.8\pm0.4\pm1.3$            & $6.7$   & $95$  & $0.1502$ \\
              & S20, $\beta=0.3$ & $10.8_{-0.4}^{+0.5}\pm2.0$   & $9.4$   & $95$  & $0.1252$ \\
              & F16              & $9.3_{-0.3}^{+0.4}\pm1.5$    & $8.2$   & $81$  & $0.1255$ \\
\hline
$\gamma$~Vel  & B15              & $6.5\pm0.1\pm0.4$            & $5.9$   & $112$ & $0.0691$ \\
              & D08              & $7.3_{-0.1}^{+0.2}\pm0.5$    & $6.8$   & $103$ & $0.0582$ \\
              & S20, $\beta=0.0$ & $6.9\pm0.1\pm0.4$            & $6.5$   & $111$ & $0.0592$ \\
              & S20, $\beta=0.1$ & $8.6\pm0.1\pm0.6$            & $7.8$   & $111$ & $0.0494$ \\
              & S20, $\beta=0.2$ & $11.6_{-0.3}^{+0.1}\pm0.8$   & $10.1$  & $112$ & $0.0394$ \\
              & S20, $\beta=0.3$ & $16.0\pm0.3\pm1.2$           & $14.3$  & $112$ & $0.0316$ \\
              & F16              & $13.7\pm0.3\pm0.8$            & $12.5$  & $94$  & $0.0463$ \\
\hline
NGC           & B15              & $22.6_{-0.7}^{+1.1}\pm1.7$   & $20.1$  & $69$  & $0.1364$ \\
2547          & D08              & $25.2\pm1.2\pm2.6$           & $22.4$  & $67$  & $0.1319$ \\
              & S20, $\beta=0.0$ & $25.6_{-1.3}^{+1.2}\pm2.2$   & $22.2$  & $69$  & $0.1128$ \\
              & S20, $\beta=0.1$ & $36.0_{-1.5}^{+1.1}\pm3.1$   & $29.9$  & $69$  & $0.0738$ \\
              & S20, $\beta=0.2$ & $47.9\pm1.4\pm4.0$   & $41.7$  & $69$  & $0.0407$ \\
              & S20, $\beta=0.3$ & $67.1_{-2.0}^{+1.8}\pm5.8$   & $61.1$  & $69$  & $0.0298$ \\
              & F16              & $44.5_{-0.3}^{+0.5}\pm2.7$   & $38.1$  & $63$  & $0.0861$ \\
\hline
NGC           & B15              & $77.3_{-1.8}^{+1.5}\pm4.5$   & $71.7$  & $215$ & $0.0638$ \\
2516          & D08              & $79.3_{-1.9}^{+1.7}\pm6.5$   & $72.7$  & $215$ & $0.0653$ \\
              & S20, $\beta=0.0$ & $78.2_{-1.3}^{+1.0}\pm6.6$   & $72.8$  & $215$ & $0.0514$ \\
              & S20, $\beta=0.1$ & $95.6_{-0.5}^{+0.7}\pm9.7$   & $88.1$  & $215$ & $0.0478$ \\
              & S20, $\beta=0.2$ & $124.3_{-0.6}^{+2.1}\pm8.1$  & $112.0$ & $215$ & $0.0554$ \\
              & S20, $\beta=0.3$ & $147.7_{-1.5}^{+3.3}\pm5.8$  & $141.3$ & $215$ & $0.0987$ \\
              & F16              & $123.6_{-0.4}^{+1.6}\pm6.5$  & $111.5$ & $146$ & $0.1220$ \\
\hline
\end{tabular}
\label{T_CMD_ages}
\end{center}
\end{table}

\subsubsection{Comparing EW(Li) with model predictions}\label{S_Analysis_EWLi}
The theoretical models provide Li depletion factors as a function of $T_{\rm eff}$ and age, which were converted to EW(Li), assuming an initial Li abundance, $A{\rm (Li)_0} = 3.3$\footnote{On the usual scale where $A{\rm (Li)} = \log (n{\rm (Li)}/n{\rm (H)}) + 12$. This is close to the Solar System meteoritic value and there is little evidence for a different value amongst the young undepleted F/G stars of many young clusters of similar metallicity to those considered here \citep[e.g.][]{2020a_Randich}.}, which is then reduced by the depletion factor and reverse-non-local thermodynamic equilibrium (NLTE)-corrected to convert it into the abundance that would be derived from a 1D atmosphere LTE curve of growth. These small correction factors are derived from the ``Breidablik'' code \citep{2013a_Magic, 2016a_Amarsi, 2021a_Wang}, which is valid for $4000<T_{\rm eff}<6000$\,K. Outside of this range the correction at the limit was used. These LTE Li abundances are then converted to EW(Li) with the curve of growth models described by \cite{2003a_Jeffries}  \citep[based on the work of][]{2002a_Zapatero-Osorio}~for $T_{\rm eff} < 4200\,$K or from \cite{1993a_Soderblom}~for $T_{\rm eff} > 4200\,$K, using the method described in \cite{2017a_Jeffries}. There will be systematic uncertainties still present in these calculations, especially for the coolest stars with $T_{\rm eff}<4000\,$K. In particular, the curves of growth described in Jeffries et al. (2003) use ``pseudo-equivalent widths" and are not calculated for $A$(Li)$<0.0$ (roughly, EW(Li)$ \simeq 150$\,m\AA). We adopt a simple extrapolation for lower abundances (where molecular contamination could be very important) to a very low $A$(Li) when EW(Li) is zero. Nevertheless, the reader should be aware that the calculation of $A$(Li) when EW(Li)$< 150$ m\AA\ and $T_{\rm eff}<4000$\,K is highly uncertain.
However, these uncertainties, particularly the contribution of molecules, would still be present if comparing (NLTE) Li abundances and $T_{\rm eff}$ with models in the theoretical plane. We prefer to compare in the observational plane so that at least the effects of observational uncertainties are apparent and almost uncorrelated on each axis.

A scatter in EW(Li) at any given spectral-type will be formed partially from a genuine dispersion but also from measurement errors. To ascertain the dominant source we compare two quantities for the cluster members in 4 spectral-type bins (excluding the strong accretors defined in $\S$\ref{S_Target_Selection}): (1) the root mean square (RMS) EW(Li) measurement error and (2) the RMS of the difference between measured EW(Li) and the interpolated EW(Li) from a linear fit with $(G-K_{\rm s})_0$ (in a given spectral-type bin). The latter quantity, denoted as $\delta$EW(Li) for each star, is indicative of the scatter in EW(Li) and is utilised in the subsequent Li-rotation analysis in $\S$\ref{S_Prot_Li}. Both quantities are presented for each cluster/spectral-type bin configuration in Table~\ref{T_EWLi_scatter}.

The EW(Li) uncertainties are broadly similar at each spectral-type but with the M2-M5 values being $\sim 1.5$ times larger, reflecting the lower SNR of these spectra. Li should be (almost) pristine in the two youngest clusters, however, the scatter for some bins is several times larger than the typical uncertainty, suggesting additional mechanisms (e.g., rotation, age spreads, undiagnosed accretion) may cause these dispersions. The  EW(Li)/$(G-K_{\rm s})_0$ distribution in NGC~2264 (top-right panel of Figures~\ref{F_SSM}~and~\ref{F_NSSM}) show almost all stars with EW(Li) $\lesssim 400\,$m\AA~are identified as strong accretors. The Li depletion appears to have initiated in the M-stars of $\gamma$ Vel (see Figures~\ref{F_SSM}~and~\ref{F_NSSM}), and the scatter is about $5-10$ times larger than the measurement error, whilst for the K-stars these values are similar. In the two oldest clusters, Li has (almost) fully depleted in the M2-M5 stars and the quantities are similar (suggesting the error bars are reasonable), whereas the scatter is several times larger than the errors for the K-stars, where Li depletion is ongoing.

{\centering
\begin{table}
\begin{center}
\begin{tabular}{p{1.35cm}p{1.25cm}p{1.25cm}p{1.25cm}p{1.25cm}}
\hline
\hline
Cluster       & K0-K5     & K5-M0     & M0-M2     & M2-M5      \\
\hline
NGC~2264      & 31.6/96.1 & 22.0/89.3  & 24.1/53.1  & 34.0/78.1 \\
$\lambda$ Ori & 33.4/30.5 & 19.3/26.0  & 23.2/41.0  & 39.8/84.2 \\
$\gamma$ Vel  & 31.0/25.3 & 20.4/55.5  & 20.4/112.6 & 28.4/241.5 \\
NGC~2547      & 28.5/81.8 & 28.0/111.8 & 20.4/62.0  & 39.9/48.2 \\
NGC~2516      & 30.8/70.2 & 21.8/58.7  & 26.1/50.3  & 48.8/62.6 \\
\hline
\end{tabular}
\caption{The root-mean-square (RMS) EW(Li) measurement uncertainties and RMS $\delta$EW(Li) values for each cluster/spectral-type bin, both in units of m\AA.}
\label{T_EWLi_scatter}
\end{center}
\end{table}}

\subsection{Comparison with standard models}\label{S_SSM}

\begin{figure*}
\centering
\begin{subfigure}
  \centering
  \includegraphics[width=.49\linewidth, height=20cm]{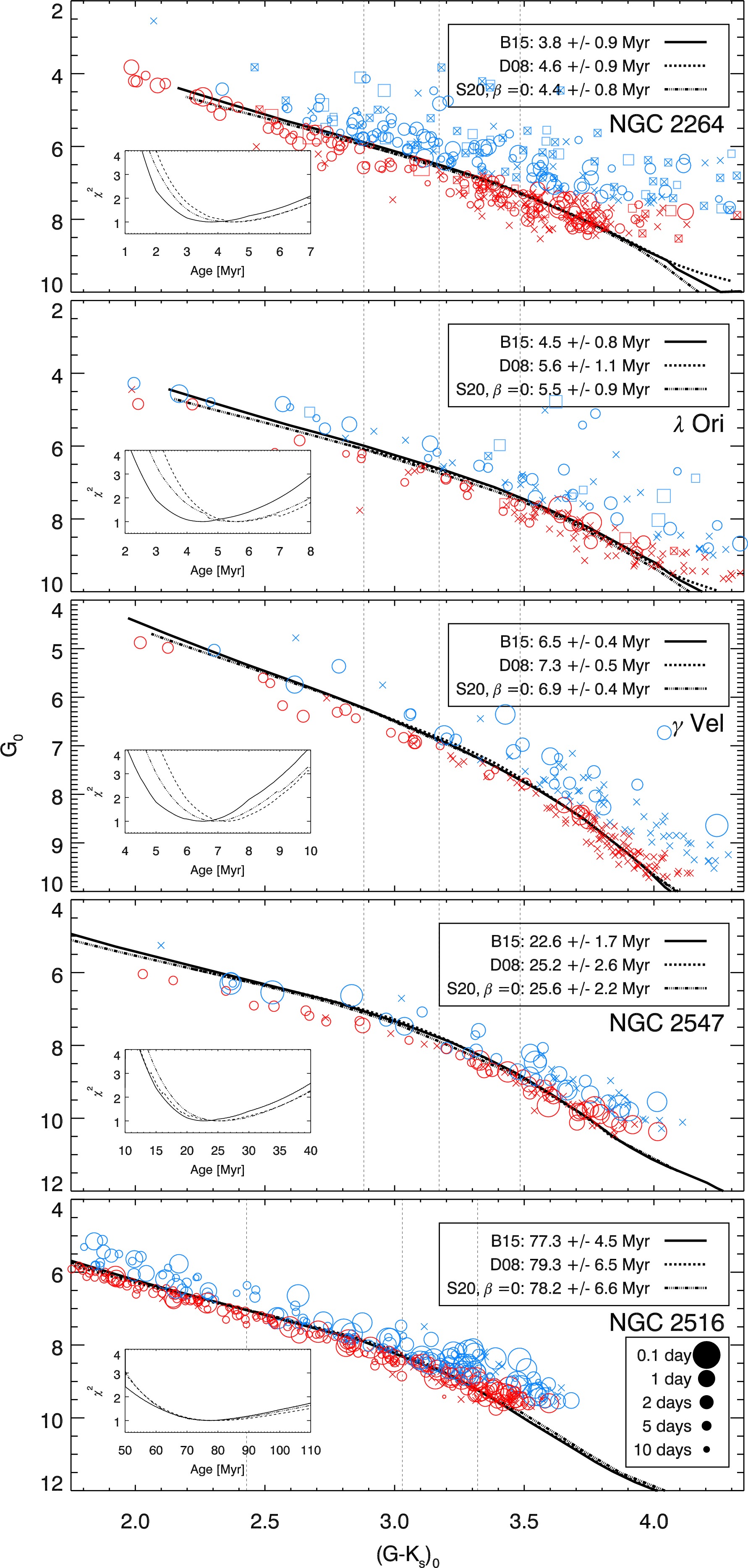}
\end{subfigure}
\begin{subfigure}
  \centering
  \includegraphics[width=.49\linewidth, height=20cm]{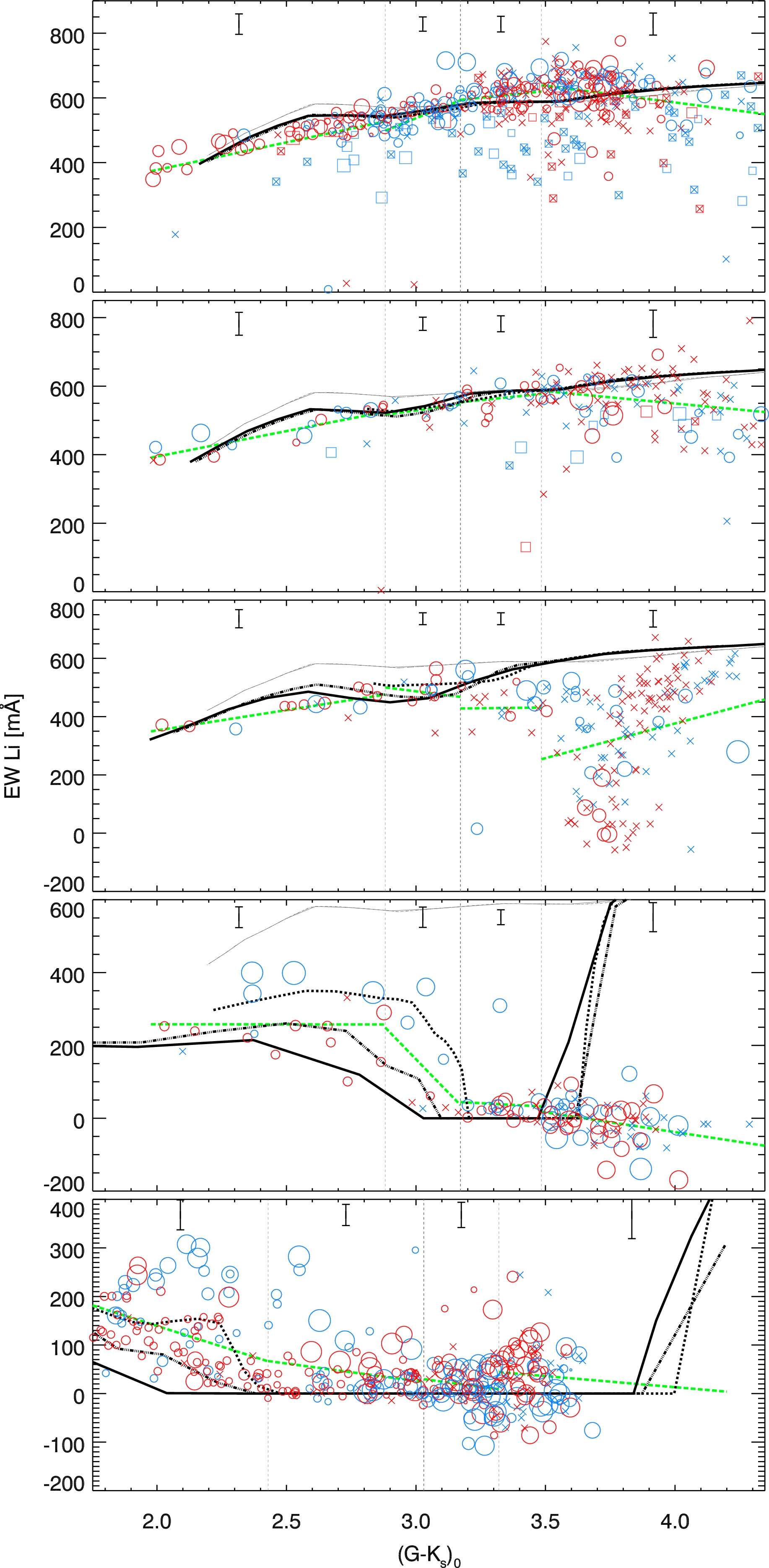}
\end{subfigure}
\caption{Left: Extinction-corrected absolute $G$ versus de-reddened $G-K_{\rm s}$ colour magnitude diagrams (CMDs) for our sample. Objects with $P_{\rm rot}$ measurements are represented with symbols scaled in size by a factor of $\log{(P_{\rm rot})}^{-1} + 1$, where circles and squares represent objects classed as weak and strong accretors, respectively. Small crosses denote objects without a $P_{\rm rot}$ measurement. Red and blue colours denote the objects used to define a single and multiple star sequence, respectively (see $\S$\ref{S_SSM}) and have the same meaning in all subsequent plots. The vertical lines represent $G-K_{\rm s}$ values corresponding to K5, M0 and M2. Solid, dotted and dot-dashed lines represent best-fit isochronal models from three evolutionary codes that invoke standard models: \citealt{2015a_Baraffe}\,(``B15''), \citealt{2008a_Dotter}\,(``D08'') and the unspotted models of \citealt{2020a_Somers}\,(``S20, $\beta=0$''), respectively. The best-fit age from each model is provided in each legend, where the error bar reflects the uncertainty in distance modulus and $E(B-V)$. The insets show the reduced $\chi^{2}$ value as a function of age close to the best-fit age. Right: EW(Li) versus $G-K_{\rm s}$ for all K- and M-type stars in our sample. Error bars in each spectral-type bin represent the median EW(Li) error and the green dashed line represents the linear fit between EW(Li) and $(G-K_{\rm s})_0$ within each spectral-type bin, as described in $\S$\ref{S_Analysis_EWLi}. The standard models corresponding to the age of the best-fit to the CMD sequence are displayed using the same symbol scheme, where the fainter  lines are the same models, but at very young ages (1.0\,Myr), representative of undepleted Li.}
\label{F_SSM}
\end{figure*}

\begin{figure*}
\centering
\begin{subfigure}
  \centering
  \includegraphics[width=.49\linewidth, height=20cm]{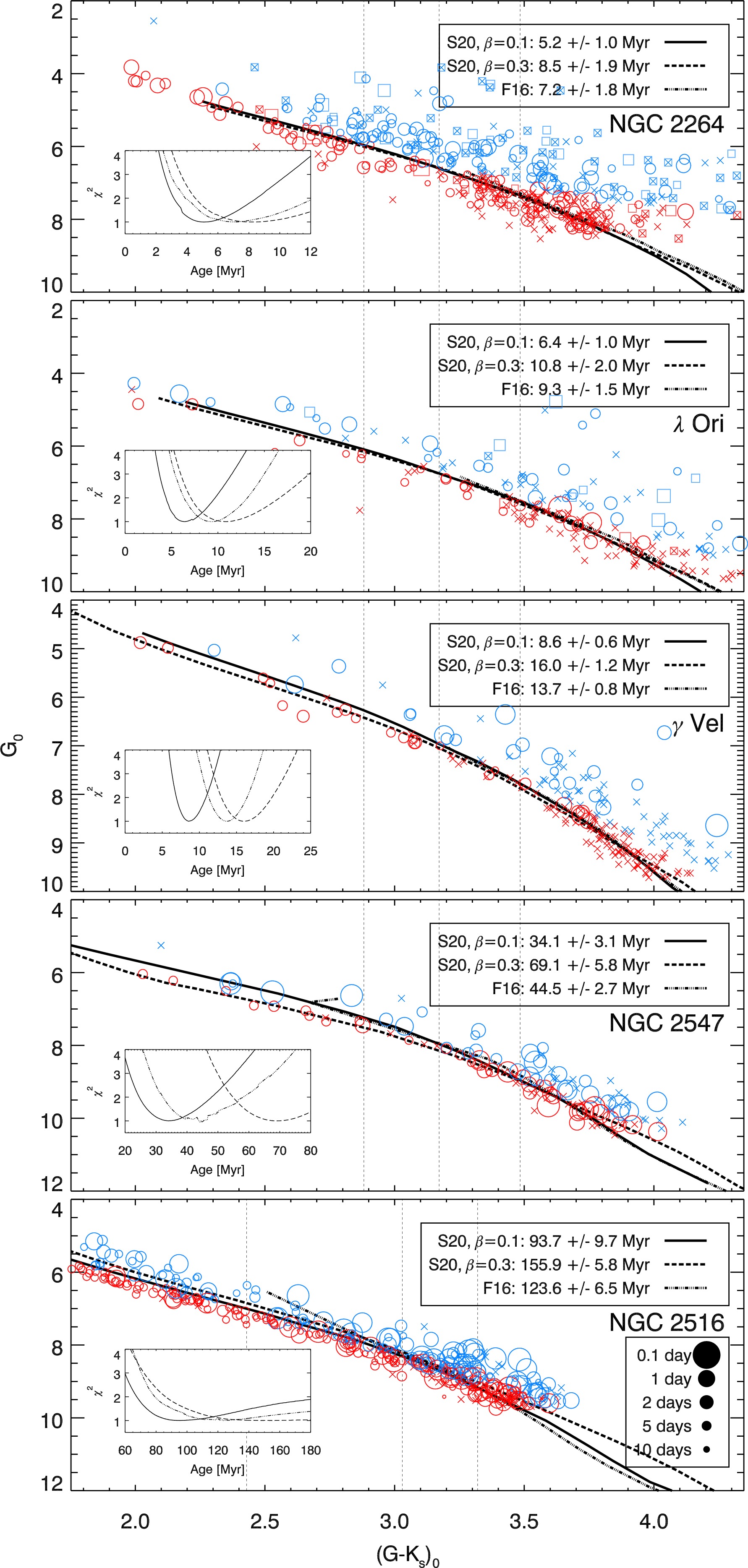}
\end{subfigure}
\begin{subfigure}
  \centering
  \includegraphics[width=.49\linewidth, height=20cm]{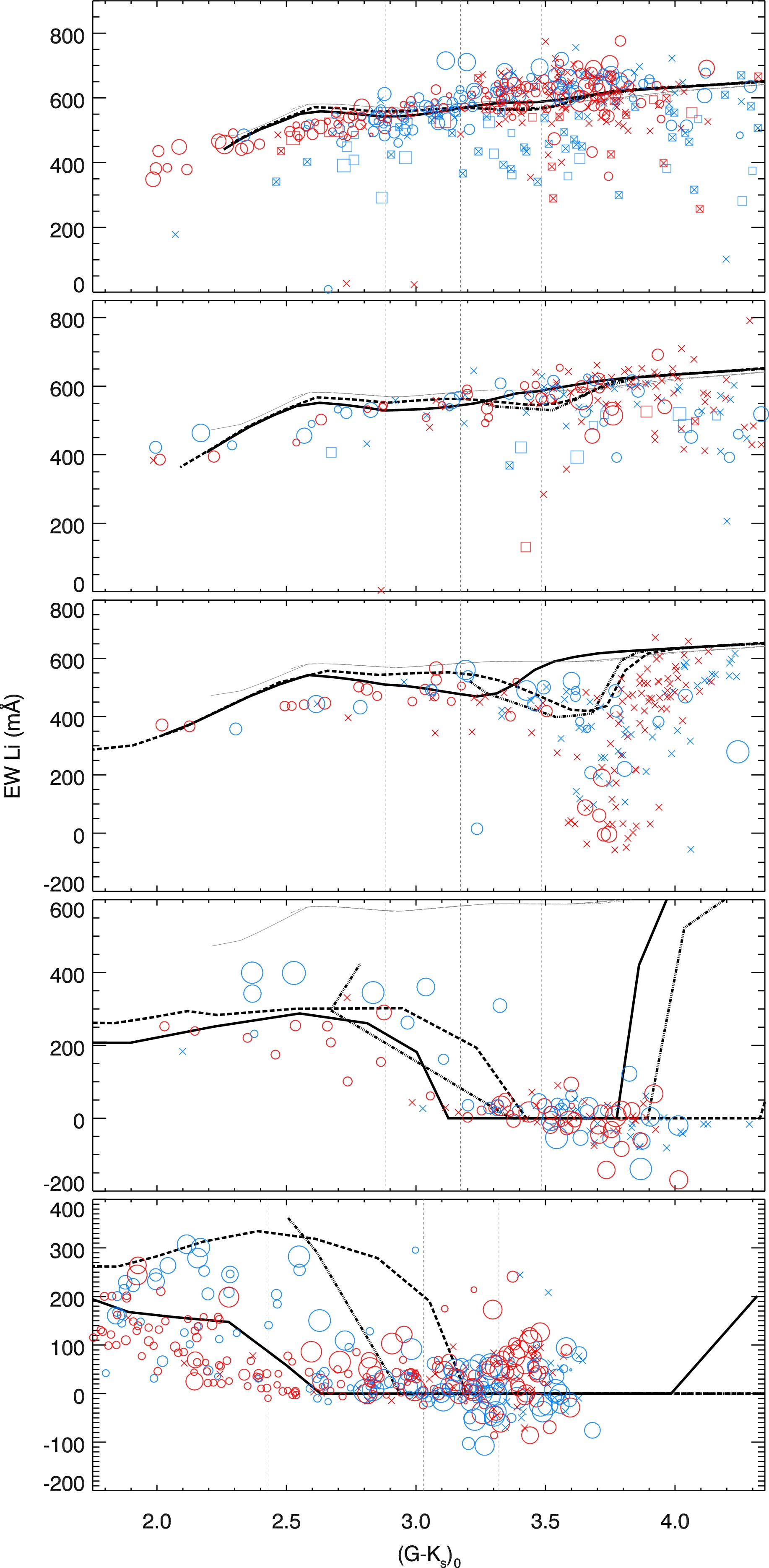}
\end{subfigure}
\caption{The equivalent data from Figure~\ref{F_SSM}, overplotted with the best-fit isochrones from the magnetic models. These are two SPOTS models fixed with $\tau=0.9$, with $\beta=0.1$~and~$0.3$ (\citealt{2020a_Somers}, ``S20, $\beta=0.1$'' and ``S20, $\beta=0.3$'') and the magnetic Dartmouth models (\citealt{2016a_Feiden}, ``F16'').}
\label{F_NSSM}
\end{figure*}

For all five clusters the CMD fits from the three standard models, shown in Figure~\ref{F_SSM} yield consistent ages (see Table~\ref{T_CMD_ages}). However the best-fitting model isochrones for the four youngest clusters do not appear to be very good fits in the CMD in the sense that the models are overluminous at higher temperatures and underluminous at lower temperatures. It would be possible to provide a good fit to the samples at $(G-K_{\rm s})_0<2.5$ using older ages, but at the expense of being underluminous by $\sim 0.3$ mag at $(G-K_{\rm s})_0\sim 3.5$ and by even more at redder colours. In contrast, the fit to the older NGC~2516 is much better over the range $1.8<(G-K_{\rm s})_0<3.5$ but the derived age of $77-81$\,Myr is about half that of the age derived from assessment of the main-sequence turn-off \citep[e.g.][gives a turn-off age of 141\,Myr for NGC~2516, compared with 100\,Myr for the Pleiades]{1993a_Meynet}.

The right-hand panels of Figure~\ref{F_SSM} show isochrones of EW(Li) versus $(G-K_{\rm s})_0$ for the same models, at the best-fitting age inferred from the CMD. The models predict that Li is undepleted in the hottest and coolest stars of the three youngest clusters. At intermediate colours the models predict a ``Li-chasm'' that grows deeper and wider for older ages. In the models, the chasm develops at $(G-K_{\rm s})_0 \sim 2.9$,  but as a relatively weak feature until at ages $>10$ Myr it deepens and broadens rapidly towards the blue. This is driven by continuing Li depletion, but also by the increasing $T_{\rm eff}$ of stars at a given mass as they develop radiative cores. In contrast, {\it the data} show a weak signature of Li depletion in $\lambda$ Ori, which becomes strikingly stronger in $\gamma$ Vel over a relatively narrow colour range from $3.5<(G-K_{\rm s})_0<4.0$, with some stars having EW(Li) consistent with total depletion. This does not match the models at all, both in terms of the amount of depletion and the colour at which it occurs. In the two older clusters there is also disagreement between models and data on the hot side of the Li-chasm, where many stars show levels of Li much higher than predicted by the models. This ``overabundance'' of Li appears strongly correlated with faster rotation rate (i.e. the larger symbols) and also with stars that are in the more luminous half of  the CMD and may be unresolved binares (the blue symbols). This is examined in more detail in $\S$\ref{S_Prot_Analysis}.

Adjusting the ages of the Li isochrones within a window that has a width of several times the error bars in age from the CMD fits does not alter these qualitative comparisons. In particular, the Li depletion that is observed in $\lambda$ Ori and $\gamma$ Vel is too strong and is at colours that are too red to be explained by the standard models {\it at any age}. 

\subsection{Comparison with magnetic models}\label{S_NSSM}

The left hand panels of Figure~\ref{F_NSSM} show fits to the CMD, for the S20 (where the fraction of photospheric flux blocked by the surrounding spots, $\beta$, is either 0.1 or 0.3) and F16 models. Best-fit ages are given in Table~\ref{T_CMD_ages} and also include results for $\beta=0.2$. These magnetic models predict best-fitting isochrones that are significantly older, by roughly a factor of two, than their non-magnetic, standard model counterparts. The S20 models produce older ages for larger $\beta$ but are roughly equivalent to the ages from the F16 models between $\beta=0.2$~and $\beta=0.3$. The reason for the older inferred ages is that the magnetic models predict that stars of a given mass and age have a larger radius (by $\sim 10-20$ per cent) and have lower $T_{\rm eff}$ than predicted by standard models.

The lower $\chi^{2}_{\rm min}$ values in Table~\ref{T_CMD_ages} indicate that the quality of the CMD fits using magnetic model isochrones is better than for the standard models (although the upper mass limit of the available F16 models prevents meaningful comparison for the warmer stars in the older clusters). A visual comparison of Figures~\ref{F_SSM} and~\ref{F_NSSM} shows that the magnetic model isochrones do better than the standard models at matching the data over the whole colour range in the four younger clusters. The S20 models with relatively large $\beta$ values are better in terms of matching the photometry of the coolest stars, though there is still a significant discrepancy for the coolest stars in the two youngest clusters.

The right hand panel of Figure~\ref{F_NSSM} shows predicted Li depletion isochrones from the same magnetic models, at the ages inferred from the CMDs. Since these are older than for the standard models in the four youngest clusters, the magnetic models predict more Li depletion despite the onset of Li depletion being delayed by radius inflation. More importantly, the $T_{\rm eff}$ and colour of the stars where Li depletion commences is shifted cooler and redward because stars of a given mass have a lower $T_{\rm eff}$ in the magnetic models. Note that this effect is independent of the assumed distance to the cluster, so can break any distance-age degeneracy from the CMD.

In NGC~2264 the models predict that Li depletion has barely begun, despite the older CMD age. This is a point we will return to in $\S$\ref{S_Prot_Li}~when considering whether there is any relationship between rotation and the apparent scatter in the observed Li depletion. The Li depletion pattern in $\lambda$ Ori is reasonably reproduced blueward of $(G-K_{\rm s})_0 = 3.0$ by all magnetic models. Redward of this point, however, only the F16 model predicts Li depletion of about the right amount, but this is at colours that are slightly too blue. 

For the older clusters, Li depletion is significant. The magnetic models appear to offer a much better description of the colours at which significant Li depletion is found in $\gamma$ Vel and potentially a much better description of the colour beyond which Li becomes completely depleted in NGC~2547 and NGC~2516.

\subsection{The cause of Li dispersion in clusters: age-spread or activity spread?}

The best-fit isochrones for the magnetic models (Figure~\ref{F_NSSM}) indicate that magnetic inhibition or greater flux-blocking by starspots leads to older inferred ages from the CMD. The Li-depletion predicted at those ages is also a better match to the colour at which depletion is seen in the older clusters. In isolation the magnetic models do not however explain why there is a significant dispersion in Li depletion observed in the M-stars of $\gamma$~Vel and the K-stars of NGC~2547 and NGC~2516. An obvious possibility is that there is a significant star-to-star variation in magnetic activity, perhaps associated with rotation rate. However, the effects of changing levels of magnetic activity or spot coverage is degenerate to some extent with age; an alternative explanation could be that an age spread within a cluster could lead to differences in Li depletion. Both possibilities could also lead to some additional dispersion in the CMD.

\subsubsection{Age spread or magnetic activity spread in $\gamma$ Vel?}\label{S_Analysis_gamVel}

Comparison of Figures~\ref{F_SSM} and~\ref{F_NSSM}~indicates that the agreement with the Li depletion data in the case of $\gamma$ Vel is more favourable for the magnetic models. The colour of the Li-dip is better reproduced by the F16 models and the S20 models, albeit still 0.1-0.2\,mag too blue. The magnetic models do predict significant Li-depletion at the best-fit isochronal age, but don't explain the many stars with EW(Li)$<300$\,m\AA\ or the few stars with very little depletion at all at $(G-K_{\rm s})_0 \sim 3.6$. 

\begin{figure}
    \centering
    \includegraphics[width=0.45\textwidth]{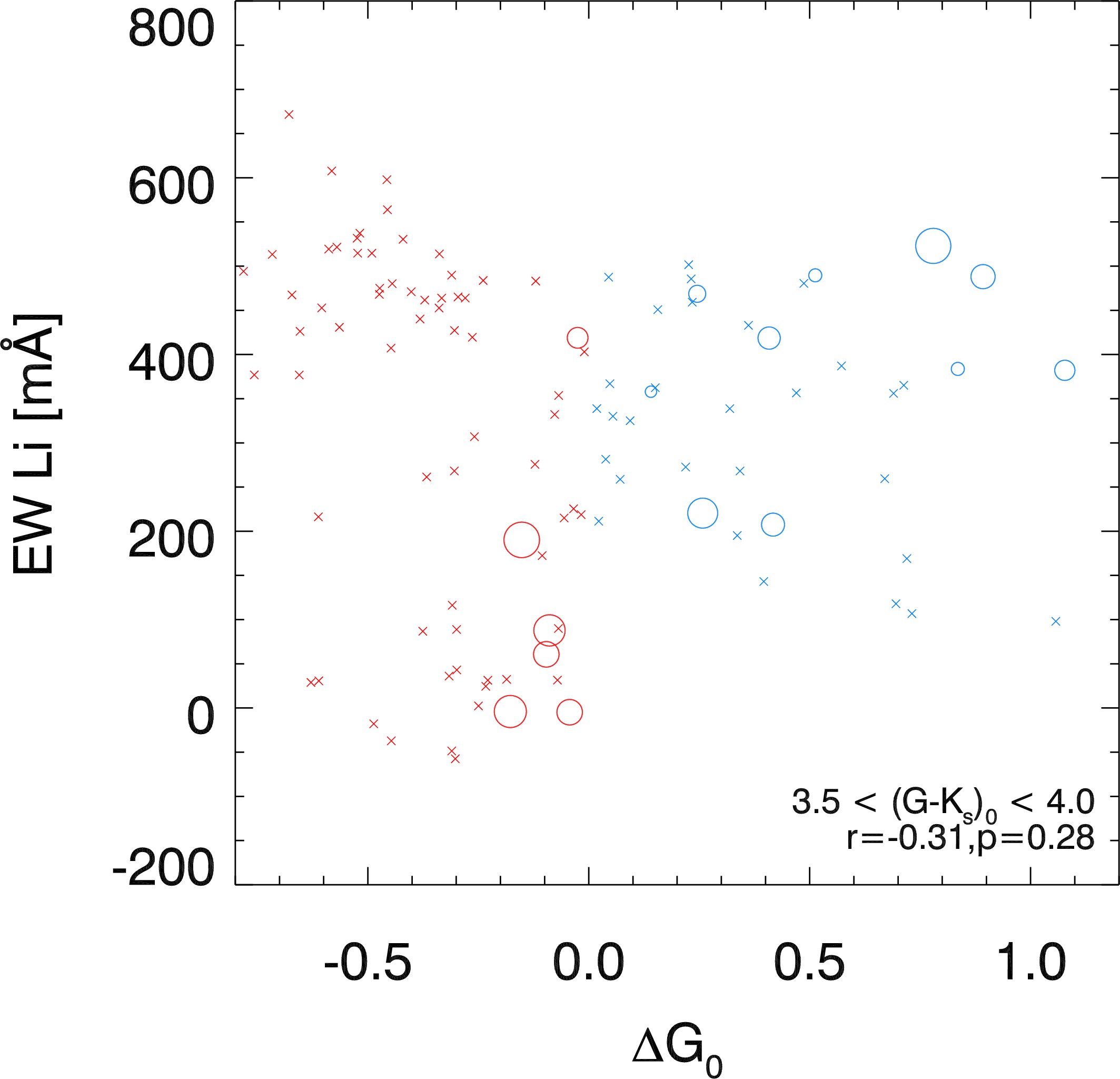}
    \caption{EW(Li) versus $\Delta G_0$ for all $\gamma$ Vel members with $3.5<(G-K_{\rm s})_0<4.0$. The colour/symbol scheme is equivalent to Figure~\ref{F_SSM}.}
    \label{F_gamVel_Gdiff}
\end{figure}

There are hints that the most Li-depleted targets are the stars we have assumed to be single. To test whether this trend is significant we select members between $3.5<(G-K_{\rm s})_0<4.0$ (i.e., corresponding to the colour range of the Li-dip) and look for any correlation between EW(Li) and the difference in absolute $G$ magnitude ($\Delta G_0$) between the observed data and that from the fit used to discriminate between bright (which could be multiple, but may be younger) and faint (likely single, but possibly older) objects at a given $(G-K_{\rm s})_0$ (see $\S$\ref{S_Analysis_CMD}). The results shown in Figure~\ref{F_gamVel_Gdiff} show a weak, negative correlation (Pearson $r$ value $=-0.31$) between EW(Li) and $\Delta G_0$ but it is clear that almost all the objects with EW(Li)$<150$\,m\AA~belong to the faint sample (i.e. red symbols). There are two obvious possible explanations for this (i) the most Li depleted objects are older or (ii) that magnetic activity inflates some stars, inhibits their Li depletion and makes them appear younger in the CMD. It is also notable that the total width of the cluster sequence in the CMD of $\sim 1.5$ mag is significantly broader than might be expected for a simple coeval population with binary systems.

\begin{figure*}
    \centering
    \includegraphics[width=0.9\textwidth]{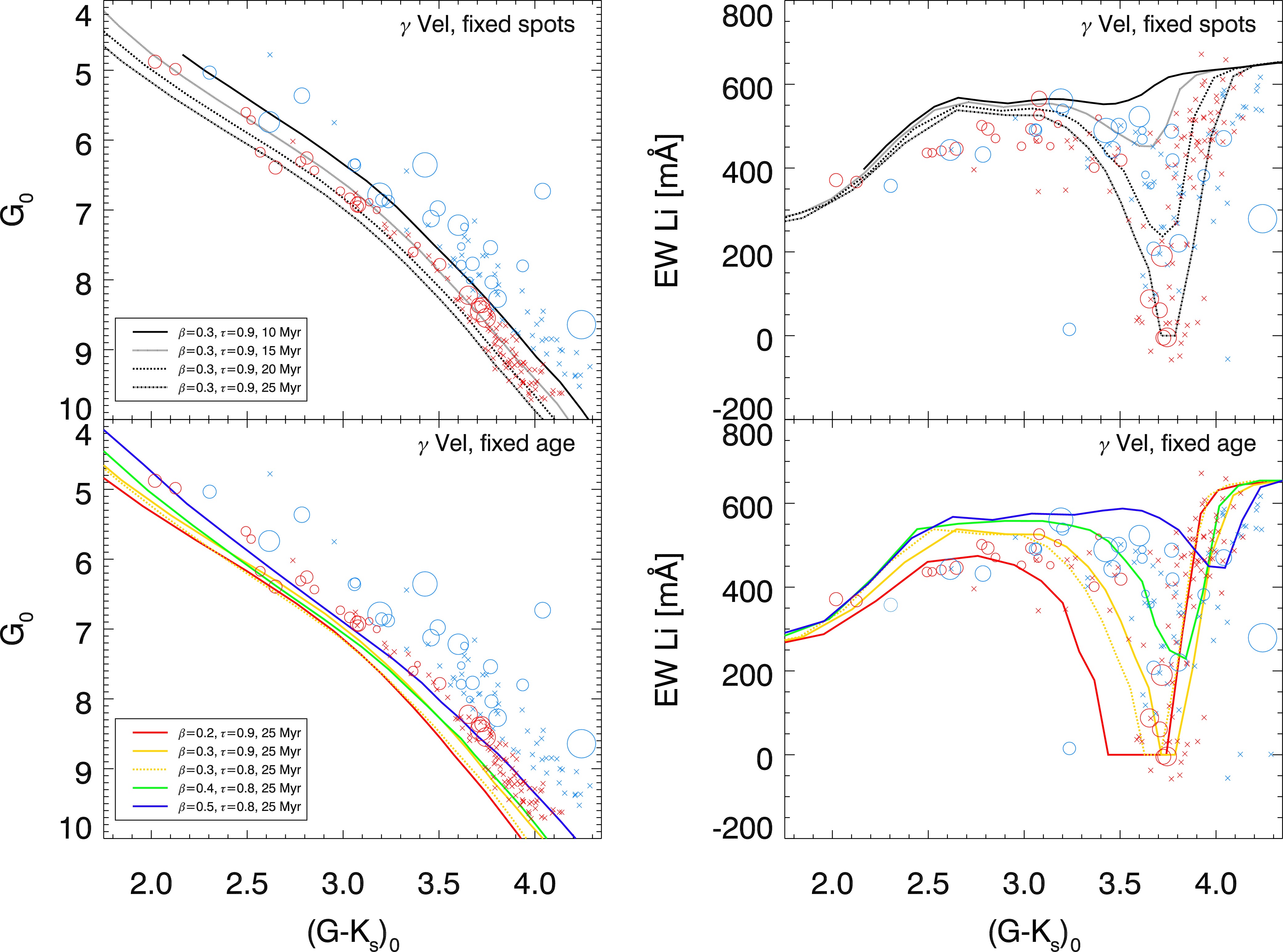}
    \caption{The $\gamma$ Vel data in the CMD (left panels) and EW(Li) versus $(G-K_{\rm s})_0$ (right panels) distribution, with S20 models overplotted for two scenarios: one set of models where the spot parameters are fixed with $\beta~=0.3$ and $\tau~=0.9$ and ages vary between 10 and 25\,Myr (top column) and another set fixed at 25\,Myr, with a variety of different spot parameters (bottom column).}
    \label{F_gamVel_Tests}
\end{figure*}

In the top two panels of Figure~\ref{F_gamVel_Tests}~we replot the $\gamma$ Vel data set with S20 models that provide plausible (visual) fits to both the CMD and Li depletion pattern, to explore whether the spread in Li depletion could be accounted for using magnetic models with a range of ages but fixed starspot parameters ($\beta=0.3$ and $\tau=0.9$). The youngest model (10\,Myr, solid black line) provides a CMD fit that looks $\sim 0.2$ mag too bright compared with the empirical single-star sequence but this could account for some of the spread in the CMD. This model also predicts very little Li depletion at all colours. In contrast, the oldest model (25\,Myr, dot-dashed line) matches the faintest stars in the CMD and also explains the stars with total Li depletion (EW(Li)$\simeq$ 0) at the correct colour. The intermediate ages of 15 and 20 Myr (thin and thick-dashed lines, respectively) both look reasonable in the CMD and encompass the majority of cluster members in and around the Li-dip.

Secondly, we investigated whether a range of magnetic activity or spot parameters could explain the Li dispersion at a fixed age. We selected 5 more models, all with ages of 25\,Myr, but explored different values of $\beta$ and $\tau$ and this set of models are plotted in the lower two panels of Figure~\ref{F_gamVel_Tests}. There is considerable degeneracy between the model age and the spot parameters: The $\beta=0.3$, $\tau=0.9$ model (solid orange line) reaches all the way down into the Li-dip, at the correct colour. Reducing the spot temperature ratio to $\tau=0.8$ at the same value of $\beta$ (thin-dashed orange line) moves the Li dip slightly blueward without changing its depth; however the CMD isochrone becomes less luminous at the same colour, and is fainter than the faintest stars. The $\beta=0.2$, $\tau=0.9$ fit (red line) predicts a Li-dip with full Li depletion, but $\sim 0.2$ mag blueward of that observed and falls well below the bottom of the data in the CMD. The $\beta=0.4$ and $\beta=0.5$ models with $\tau=0.8$ (green and blue lines, respectively) predict progressively less Li depletion as the spot coverage grows and also sit at higher luminosities in the CMD and might explain the less Li-depleted stars.

In summary, it seems it would take an age spread of $>10$ Myr to explain the full range of Li depletion if this were the only factor. We cannot rule this out, though it should be noted that the $\gamma$~Vel sample is restricted to the high probability members of ``population A" \citep[as defined by][]{2014b_Jeffries}, which is the population most tightly clustered around the star $\gamma^2$ Vel, and does not include ``population B", which is a more dispersed population found over the broader Vela OB2 region. An alternative explanation could be a range of magnetic activity and spot properties. A range of $\beta$ and $\tau$ at fixed age seems almost capable of explaining the full range of Li depletion at a fixed age but may struggle to explain the width of the cluster sequence in the CMD. It is of course possible that a combination of a (smaller) age spread and a dispersion of magnetic activity is responsible.

\subsubsection{An age spread or magnetic activity spread in NGC~2547 and NGC~2516?}\label{S_Analysis_Older}

\begin{figure*}
    \centering
    \includegraphics[width=0.9\textwidth]{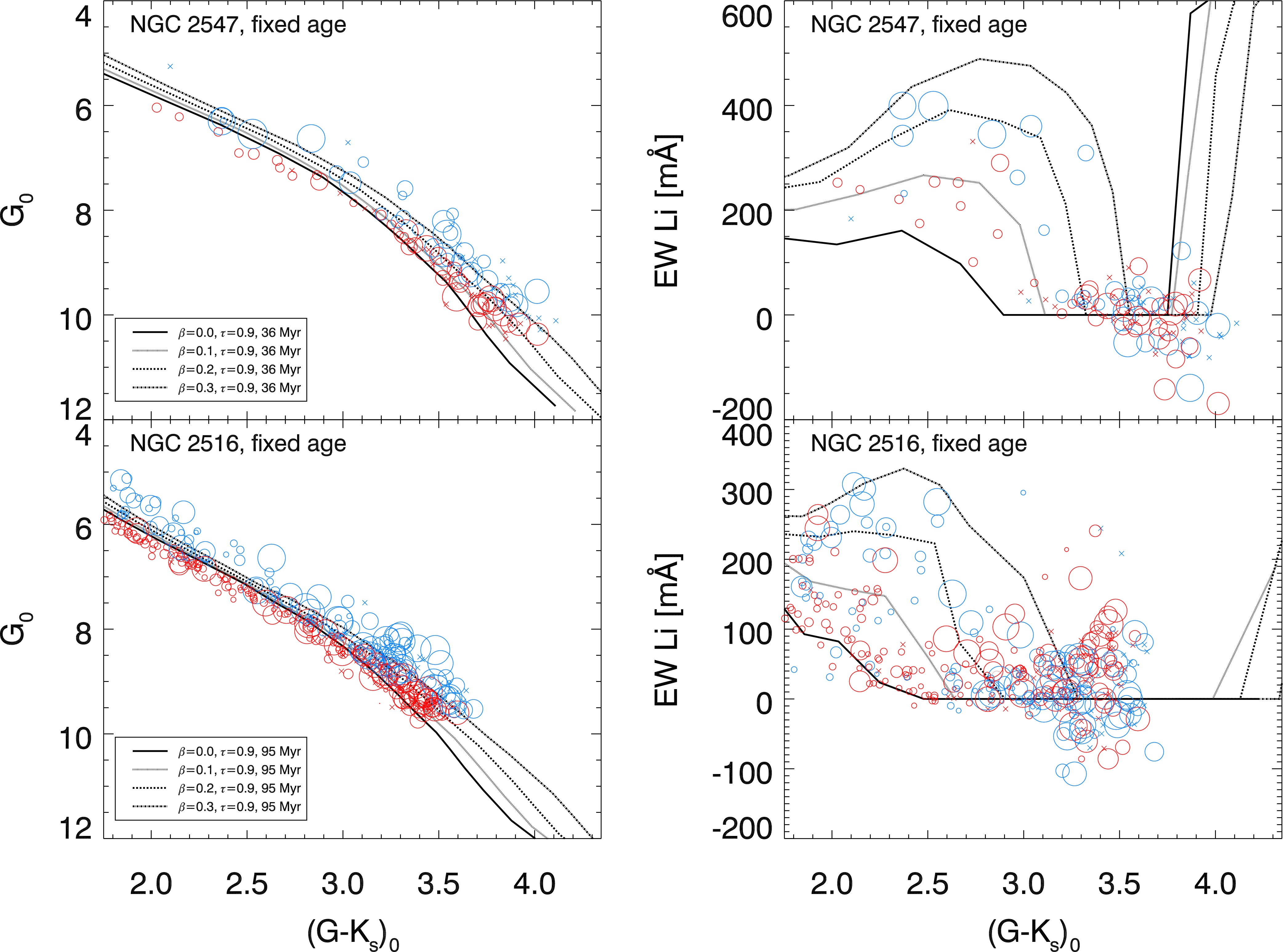}
    \caption{The NGC~2547 and NGC~2516 data (top and bottom row, respectively) in the CMD (left panels) and EW(Li) versus $(G-K_{\rm s})_0$ (right panels) distribution, with S20 models fixed at 36 and 95\,Myr, respectively, in the range $0 \leq \beta \leq 0.3$.}
    \label{F_2547_Tests}
\end{figure*}

In NGC~2547 and NGC~2516 the dispersion in Li depletion is most prominent in the K-stars. In both the standard and magnetic models, at the best-fit CMD ages, such stars have almost reached the ZAMS and PMS Li depletion should have been halted by the formation of a radiative core. This means that, in contrast to $\gamma$~Vel, the effects of a modest age dispersion ($\sim 10$ Myr) are much smaller in terms of broadening the CMD or producing differences in Li depletion once stars have reached the ZAMS. It is notable that the CMD cluster sequences for NGC~2547 and NGC~2516 are much narrower than for $\gamma$~Vel or the younger clusters. 

A hypothesis that has been made in the past \citep[e.g.][]{2018a_Bouvier, 2021a_Jeffries} is that the dispersion in Li depletion among the K-stars could be due to a spread of magnetic activity that is correlated with the (present) rotation rates of the stars. Figure~\ref{F_2547_Tests} shows a set of spot models with different values of $\beta$ but at a fixed age (using the best-fit CMD age for $\beta=0.1$) and $\tau$. These do indeed show that by varying the spot coverage between $0<\beta<0.3$, the full range of Li depletion in the K- and early-M stars of these two clusters can be explained, with higher levels of spot coverage leading to less Li depletion. Furthermore, all of these models sit well within the cluster sequences in the CMD, with a prediction that the more spotted and most Li-rich stars would have a higher luminosity at any given colour. 

\section{Rotational Trends}
\label{S_Prot_Analysis}

\begin{figure}
    \centering
    \includegraphics[width=0.45\textwidth]{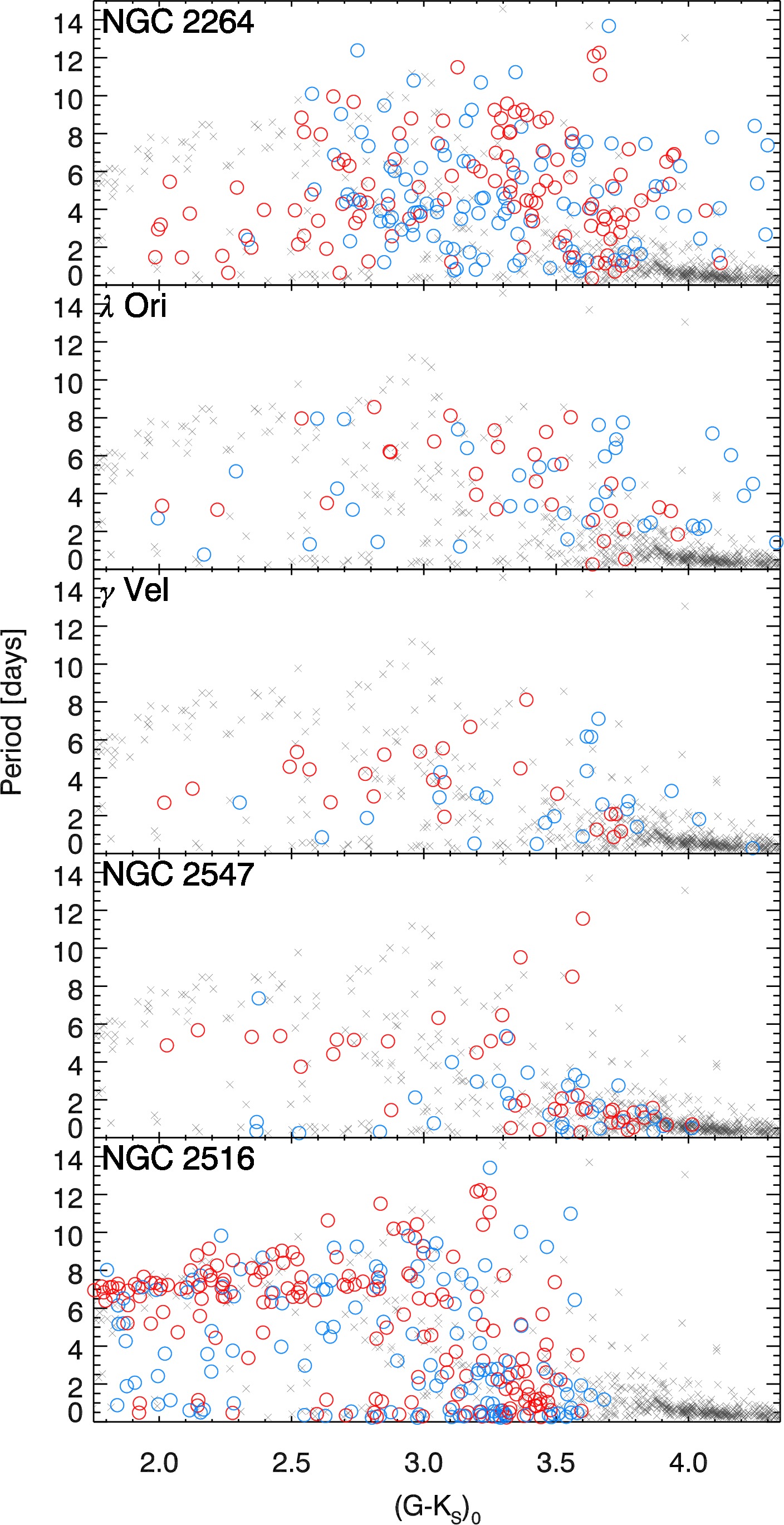}
    \caption{Plots of $P_{\rm rot}$ versus $(G-K_{\rm s})_0$ for the 5 clusters analysed in this work. Grey crosses denote equivalent data from a sample of Pleiades members observed during the Kepler-K2 campaign \citep{2016a_Rebull}. Red and blue symbols correspond to the faint and bright samples defined in $\S$\ref{S_Analysis_CMD}.}
    \label{F_Period_GKs}
\end{figure}

Figure~\ref{F_Period_GKs} shows $P_{\rm rot}$ versus $(G-K_{\rm s})_0$ for all cluster targets with a period measurement in Table~\ref{T_Periods}. The general features of the distributions and their progression with time agree with empirical features seen in the framework of ``Gyrochronology" \citep{2003a_Barnes,2014a_Bouvier}. In the 3 youngest clusters, at any given $(G-K_{\rm s})_0$, there is an order-of-magnitude spread in $P_{\rm rot}$ (red/blue open circles in Figure~\ref{F_Period_GKs}, panels 1--3). These stars then typically spin-up by a factor of $\sim 2$ at the age of $\gamma$~Vel. In the 2 older clusters the slowest rotators become slower and the fastest rotators have become faster, such that there are signs of an I- and C-sequence developing \citep[see][]{2003a_Barnes}, where the warmer stars have an approximately bimodal period distribution. The brightest stars (blue symbols) tend to rotate faster on average than the fainter stars (red symbols) in the older clusters -- a feature that has been noted in the Pleiades \citep[e.g.,][]{2016a_Stauffer}. As a  comparison, a sample of Pleiades stars in the same colour range is shown \citep[][grey crosses in Figure~\ref{F_Period_GKs}]{2016a_Rebull}. The NGC~2516 stars, that have a similar age to the Pleiades, closely follow the Pleiades data \citep[see also][]{2020a_Fritzewski}. 

In $\S$\ref{S_Analysis}~we showed that models incorporating magnetic fields or starspots could better describe the CMD and overall Li depletion pattern of this collection of young clusters. Since there is a significant dispersion in rotation rates in all these clusters, then there might also be a dispersion in magnetic properties that leads to correlations between rotation rates and position in the CMD and Li depletion \citep[see also][]{2016a_Covey, 2017a_Somers, 2021a_Jeffries}.
In both the CMDs and EW(Li) plots (Figures~\ref{F_SSM}~and~\ref{F_NSSM}), the objects with $P_{\rm rot}$ measurements are scaled in size by a factor of $\log(P_{\rm rot}/{\rm d})^{-1} + 1$, such that rapid and slow rotators are denoted by large and small symbols, respectively. To investigate how CMD position and Li-depletion depend on rotation and spectral-type we divide members of each cluster into 6 spectral-type bins: all K-stars; K0-K5; K5-M0; all M-stars; M0-M2 and M2-M5. For some cluster/spectral-type configurations there are not enough data to provide a meaningful analysis (usually due to a lack of $P_{\rm rot}$ measurements), but where enough measurements are available for a given cluster, we investigate the trends of rotation with CMD position ($\S$\ref{S_Prot_CMD}) and Li depletion ($\S$\ref{S_Prot_Li}). 

In order to check the strongest possible effects of potentially unreliable $P_{\rm rot}$ values, we tried experiments where a random 16 per cent (see \S\ref{S_Prot_Overview}) of the periods were replaced with values sampled randomly from the entire cluster. As expected, this does slightly weaken the p-values of strong correlations but does not introduce significant correlations where none were originally present. This suggests that where we report a correlation is it likely that any period unreliability has acted to weaken that correlation and it is probably more significant.

\subsection{Is CMD position related to rotation?}\label{S_Prot_CMD}

\begin{figure*}
    \centering
    \includegraphics[width=0.9\textwidth]{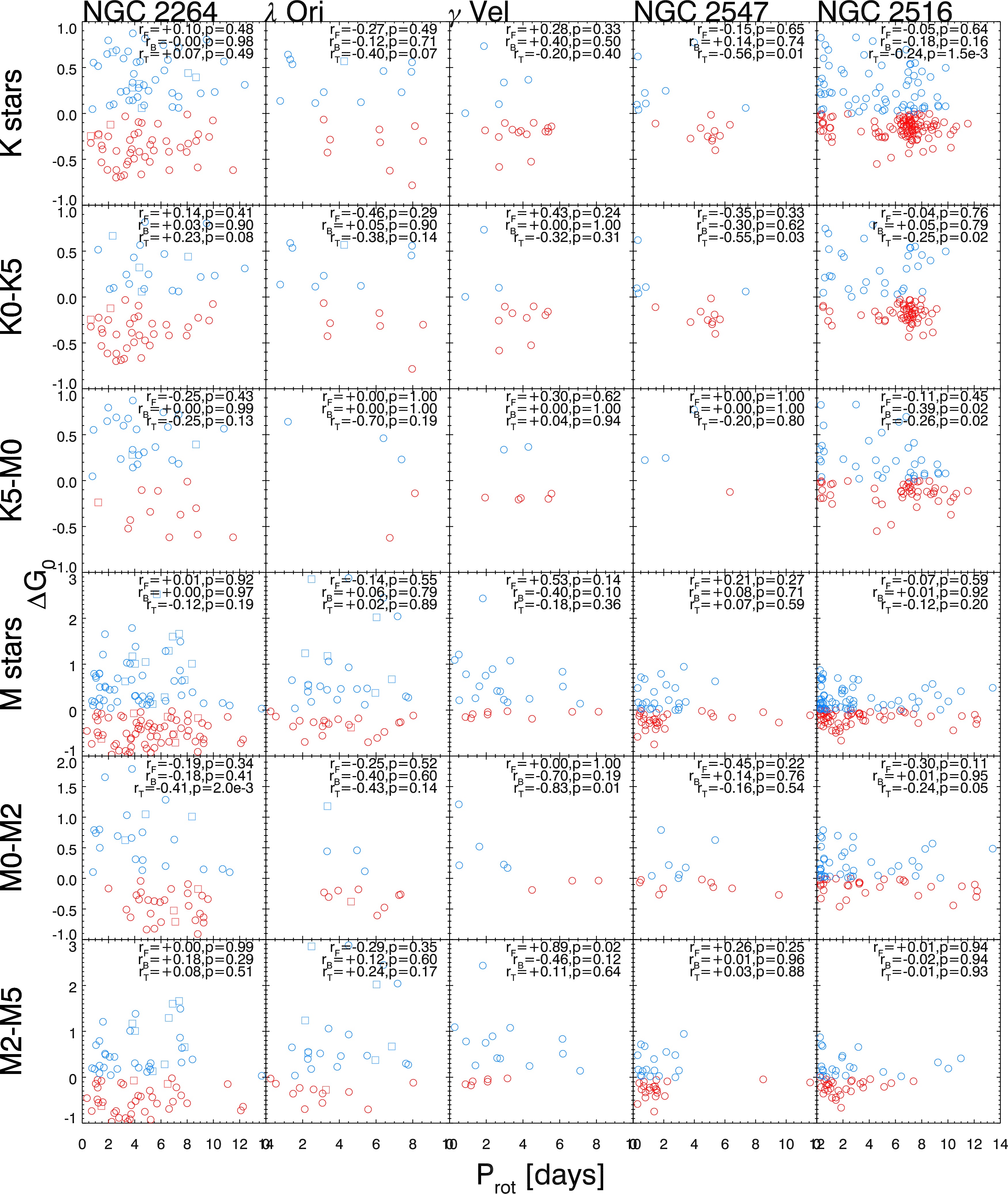}
    \caption{$\Delta G_0$ versus rotation period ($P_{\rm rot}$) for the 5 clusters considered in this work, separated into their 6 spectral-type bins (see $\S$\ref{S_Prot_CMD}). Blue and red symbols denote bright and faint objects, respectively. The values in the upper-right panels of each sub-plot are the Pearson $r$ correlation coefficient and the corresponding p-value for the single, binary and combined sample, respectively.}
    \label{F_Prot_Gdiff}
\end{figure*}

Figure~\ref{F_Prot_Gdiff}~shows the displacement from the best-fit isochrone, $\Delta G_{0}$ (see $\S$\ref{S_Analysis_gamVel}, where a positive value is brighter than the isochrone) versus $P_{\rm rot}$ for each cluster/spectral-type configuration, where red and blue stars denote faint and bright stars, respectively. We calculate the Pearson $r$ correlation coefficient and the corresponding $p$-value for the faint objects ($r_{\rm F}$), bright objects ($r_{\rm B}$) and the combined population ($r_{\rm T}$), as long as there are $>3$ objects in the sample. These are provided in the top-right corner of each panel of Figure~\ref{F_Prot_Gdiff}. Here we regard $p$-values $<\,0.05$ as statistically significant results. Weak, moderate and strong correlations are regarded as having $|r_{\rm T}| <0.3$, $0.3 \leq |r_{\rm T}|< 0.5$ and $|r_{\rm T}|  \geq 0.5$ respectively. 

\paragraph*{K-stars:} There are no significant trends among the K-stars (or the K0-K5 and K5-M0 subsamples) in NGC~2264, $\lambda$~Ori or $\gamma$~Vel. The results for NGC~2264 is notable in the sense that there are a large number of rotation periods with which to find any correlation. There are however signifcant strong and negative correlations in the K-stars and K0-K5 stars of NGC~2547 (in the sense that stars with short rotation periods appear to be more luminous) and weak and negative, but still highly significant, correlations in the K-type, K0-K5 and K5-M0 stars of NGC~2516.

\paragraph*{M-stars:} Across the whole M-star range, correlations are generally weak and with little significance for all clusters. However, the M0-M2 subsamples combining both bright and faint stars show increasingly strong negative trends in NGC~2264, $\lambda$~Ori and $\gamma$~Vel (that are statistically significant for NGC~2264 and $\gamma$~Vel), but becoming weaker in the two older clusters (but still significant in NGC~2516). No significant trends are present among the M2-M5 stars of any of the clusters.

\subsection{Li and rotation}\label{S_Prot_Li}

\begin{figure*}
    \centering
    \includegraphics[width=0.9\textwidth]{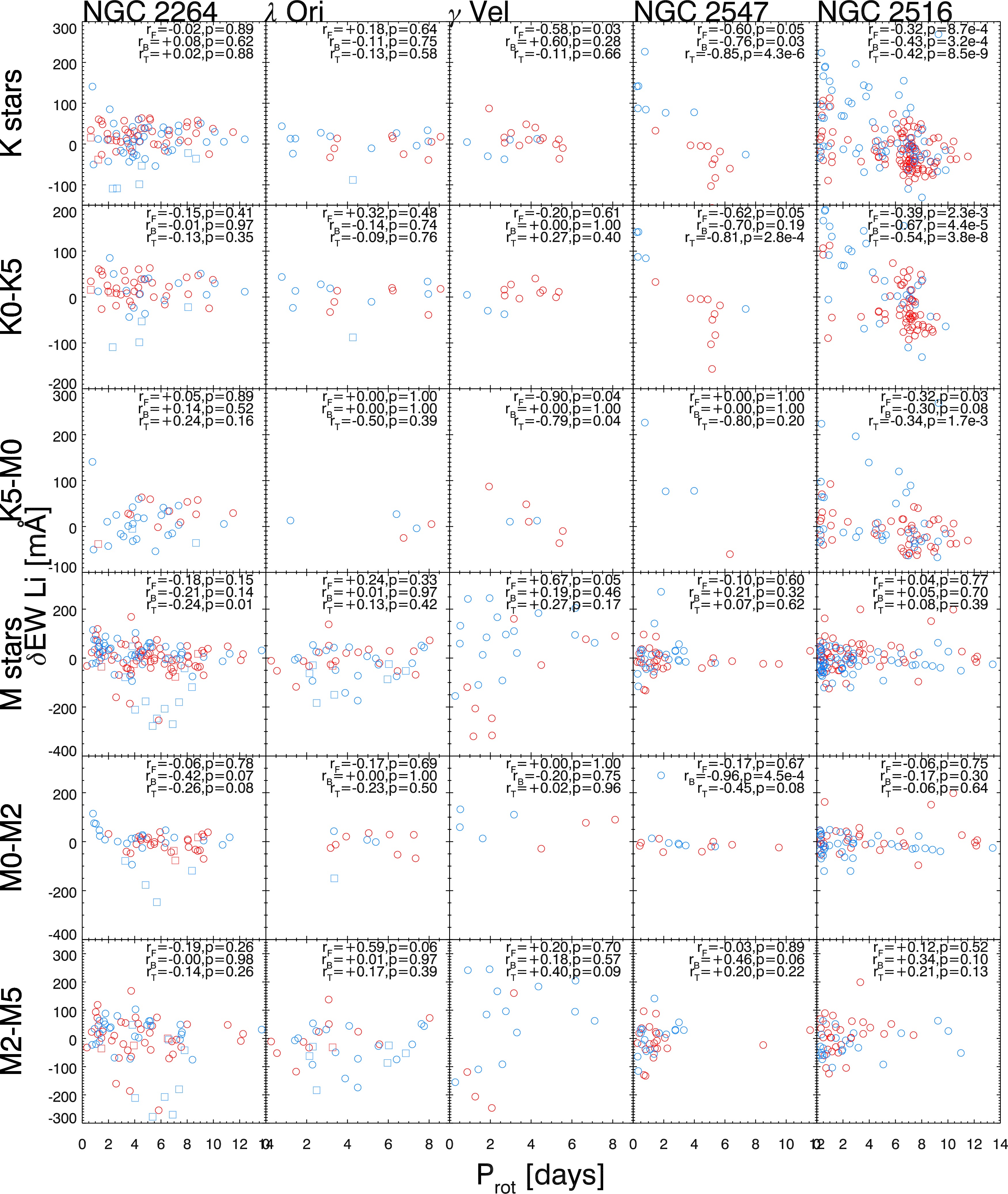}
    \caption{Rotation period ($P_{\rm rot}$) versus $\delta$EW(Li). The symbol scheme and text in the top-right corner of each panel are equivalent to the those described in the caption of Figure~\ref{F_Prot_Gdiff}.}
    \label{F_Prot_EWdiffN}
\end{figure*}

Using methods similar to those described in $\S$\ref{S_Prot_CMD}, we investigate trends between $P_{\rm rot}$ and Li-depletion. The quantity used to describe the relative amount of Li-depletion in each cluster is $\delta$EW(Li), as defined in $\S$\ref{S_Analysis_EWLi}. Figure~\ref{F_Prot_EWdiffN}, which shows the $P_{\rm rot}/\delta$EW(Li) distributions for each cluster/spectral-type range, has the same format as Figure~\ref{F_Prot_Gdiff}.

For most of the panels in Figure~\ref{F_Prot_EWdiffN} the $P_{\rm rot}/\delta$EW(Li) distributions of the faint and bright stars are not significantly different, however there are indications among the K-stars in the two oldest clusters and the M2-M5 stars of $\gamma$~Vel that the brighter objects tend to be less Li-depleted. The impact of brightness and binarity on Li-depletion is discussed further in $\S$\ref{S_Discussion_Rotation}. In the following we assess the $P_{\rm rot}/\delta$EW(Li) trends for the combined samples only, where positive/negative correlations refer to slower/faster rotators being less Li-depleted.\\

\paragraph*{K-stars:} There are no significant correlations among the K-stars of the two youngest clusters, but a strong, just significant (based on only 7 stars) negative correlation in the K5-M0 stars of $\gamma$ Vel.

The two older clusters exhibit very strong negative correlations in their K-stars that are highly significant. This is the ``classical" Li-rotation relationship reported in the Pleiades and many other clusters (see \S\ref{S_Introduction}). In NGC~2516, the correlation is much stronger in the K0-K5 stars than the K5-M0 stars, which mostly have EW(Li) consistent with zero, but there are a small group of bright stars that are both rapidly rotating and (relatively) Li-rich.\\

\paragraph*{M-stars:} There are no strong/significant trends in any of the M-star samples other than a significant, weak, negative correlation in the M-stars of NGC~2264. That no correlations are seen in the oldest two clusters is unsurprising, since all these stars have likely (almost) fully depleted their Li. It is notable that the wide dispersion in $\delta$EW(Li) seen in the M2-M5 stars of $\gamma$~Vel appears not to be negatively correlated with rotation. If anything there is a marginal, moderate, positive correlation (i.e. the slower rotators have more Li). \\

Our results offer mixed support for the findings of \citet[][herein, B16]{2016a_Bouvier}. B16 found a weak but significant correlation between $\delta$EW(Li) (calculated in much the same way as here, but based on an earlier GESiDR4 dataset and using different EW(Li) estimation techniques) and rotation period for NGC~2264 stars in the range K4-M2 approximately. We have compared our results against B16 in Appendix~\ref{S_Bouvier16}. We find no correlation between rotation and $\delta$EW(Li) in this spectral type range based on our larger NGC~2264 dataset. The difference is probably attributable to differences in EW(Li) estimation associated with blending and continuum estimation in fast rotators but we do find a correlation in the M-stars, which may chiefly be driven by stars in the M0-M2 subsample, which show a marginally significant correlation.

\section{Discussion}\label{S_Discussion}

\subsection{Implications for stellar models}\label{S_Discussion_Models}

In \S\ref{S_NSSM} it was shown that magnetic models provide improved (although not perfect) fits to both the CMD and EW(Li)/colour diagram and predict older ages compared to their standard model counterparts. This has now been noted in several clusters in the age range of those studied in this work, e.g., Sco-Cen \citep[$\sim 10\,$Myr,][]{2016a_Feiden}, $\gamma$ Vel \citep[18-21\,Myr,][]{2017a_Jeffries} and NGC~2232 \citep[38\,Myr,][]{2021a_Binks}. By combining data from homogeneous sources, this work provides an important step to quantifying these comparisons by reducing systematic biases and testing the models across a range of ages and masses. Our result agree with previous claims that cluster ages derived from standard isochronal fits to low-mass stars in young clusters may be underestimated by a factor of two in the youngest clusters considered here \citep{2013a_Bell}.

The age discrimination of the isochronal fits in the CMD arises primarily whilst stars are in the PMS phase. Once stars reach the ZAMS, which occurs at higher masses first, age sensitivity diminishes and the statistical age uncertainties increase. In addition, there are systematic uncertainties arising from the degeneracy between age and level of magnetic activity/spot coverage in the CMD, in the sense that models with higher levels of spot coverage lead to much older inferred ages (see Table~\ref{T_CMD_ages}). This degeneracy is broken to some extent by including the information provided by Li depletion. In all the clusters, and colour ranges within those clusters where Li depletion is observed, the older ages inferred using magnetic models in the CMD provide better predictions of that Li depletion, both in terms of the extent of Li depletion and the colour where it is observed.

It is possible that the stars in these clusters have a range of magnetic activity and spot parameters or that levels of activity and spottedness may vary systematically with mass. For example, using spectroscopic diagnostics, \cite{2016a_Fang}~measure a range of spot coverage in the K/M-stars of the Pleiades, ranging from almost spotless to $\sim$50 per cent spot fractions. Isochrones where the M-dwarfs were more spotted than the K-dwarfs would likely provide an improved fit to the CMD in all of the clusters. This might be plausible on the basis that they have deeper convection and smaller Rossby numbers for the same rotation period \citep[e.g.,][]{2011b_Jeffries, 2011a_Wright}. A spread in magnetic activity or in the spot temperatures at the same observed colour could also help to explain the dispersion in Li depletion that is clearly seen in $\gamma$ Vel and the older clusters. The Li-depletion patterns of the K-stars in NGC~2547 and NGC~2516 are reasonably encompassed by S20 isochrones that span a range of spot flux-blocking factors from zero to $\beta=0.3$ (the latter corresponding to 87 per cent of the surface being covered by spots with $\tau=0.9$, see Figure~\ref{F_2547_Tests}). If this were the case then a dependence of magnetic activity on rotation might then also explain the strong correlations between Li-depletion and rotation and luminosity in the K-stars of these clusters (see \S\ref{S_Discussion_Rotation}). 
In NGC~2547, the CMD ages of $36.0 \pm 3.4$ Myr inferred from the $\beta=0.1$ spotted model and $44.8 \pm 0.2$ Myr from the Feiden magnetic models are in much better agreement with the significantly less model-dependent Li Depletion Boundary age of $35 \pm 3$ Myr found by \cite{2005a_Jeffries} \citep[which would be revised slightly upwards to 37\,Myr with the addition of starspots][]{2014a_Jackson} than the CMD age from the standard isochrones of 24--27\,Myr. A similar conclusion was reached for the young cluster NGC~2232 by \cite{2021a_Binks}.

Models that incorporate starspots have a natural degeneracy between the amount of flux blocked by the spots ($\beta$) and the temperature ratio of the spot and photosphere ($\tau$). For a given value of $\beta$, larger values of $\tau$ generally lead to older CMD fits because the isochrones shift redward. In the case of $\gamma$ Vel (\S\ref{S_Analysis_gamVel}), we tuned the spot parameters at a fixed age and found a set of reasonable CMD fits that were capable of encompassing the range of Li depletion seen in its M-stars (see Figure~\ref{F_gamVel_Gdiff}). However, there is still a degeneracy with age in the sense that the range of Li depletion could also be explained with age spreads that are $\geq 10$ Myr. Indeed, some age spread seems to be required in order to explain the width of the cluster sequence in the CMD. The same degeneracy is less problematic in the older clusters because $\sim 10$ Myr age spreads have almost no effect on the predicted Li depletion or CMD position once stars have reached the ZAMS.

The cluster dataset used here overlaps with the GES dataset considered by \cite{2022a_Franciosini} and our findings are in partial agreement. Franciosini et al. also conclude that spotted models with $\beta \sim 0.2$ are better fits to the CMD and Li depletion pattern of $\gamma$~Vel population A and at an older age than would be provided by standard models. They also find that the CMD and Li-depletion patterns of NGC~2547 and NGC~2516 could be fitted by unspotted models, if considering only stars with $T_{\rm eff}>4500$\,K, but that spotted models provide a much better representation of the cooler objects. In contrast we find that the spotted models unambiguously work better than the unspotted models, possibly because we have restricted our analysis to star cooler than spectral type K0, which is where the majority of any age-sensitivity arises in both diagrams.

The older ages inferred from magnetic models have obvious implications for the empirical estimation of timescales associated with the evolution of PMS stars and their circumstellar environments, which are usually calibrated using groups of stars with assigned isochronal ages. These include the timescales associated with accretion processes for the youngest clusters \citep{2020a_Bonito}, the lifetimes of protoplanetary disks \citep{2006a_Carpenter,2015a_Ribas,2016a_Li}~dissipation rates of debris disks \citep{2017a_Binks,2020a_Silverberg}~and the angular momentum evolution for Solar-type and low-mass stars in clusters \citep{2013a_Gallet,2019a_Amard}. The clusters studied in this work are at the forefront of near-future exoplanet surveys \citep{2019a_Kastner}, allowing astronomers to empirically test models of planetary formation and evolution in the first $\sim 100\,$Myr.

\begin{figure}
    \centering
    \includegraphics[width=0.45\textwidth]{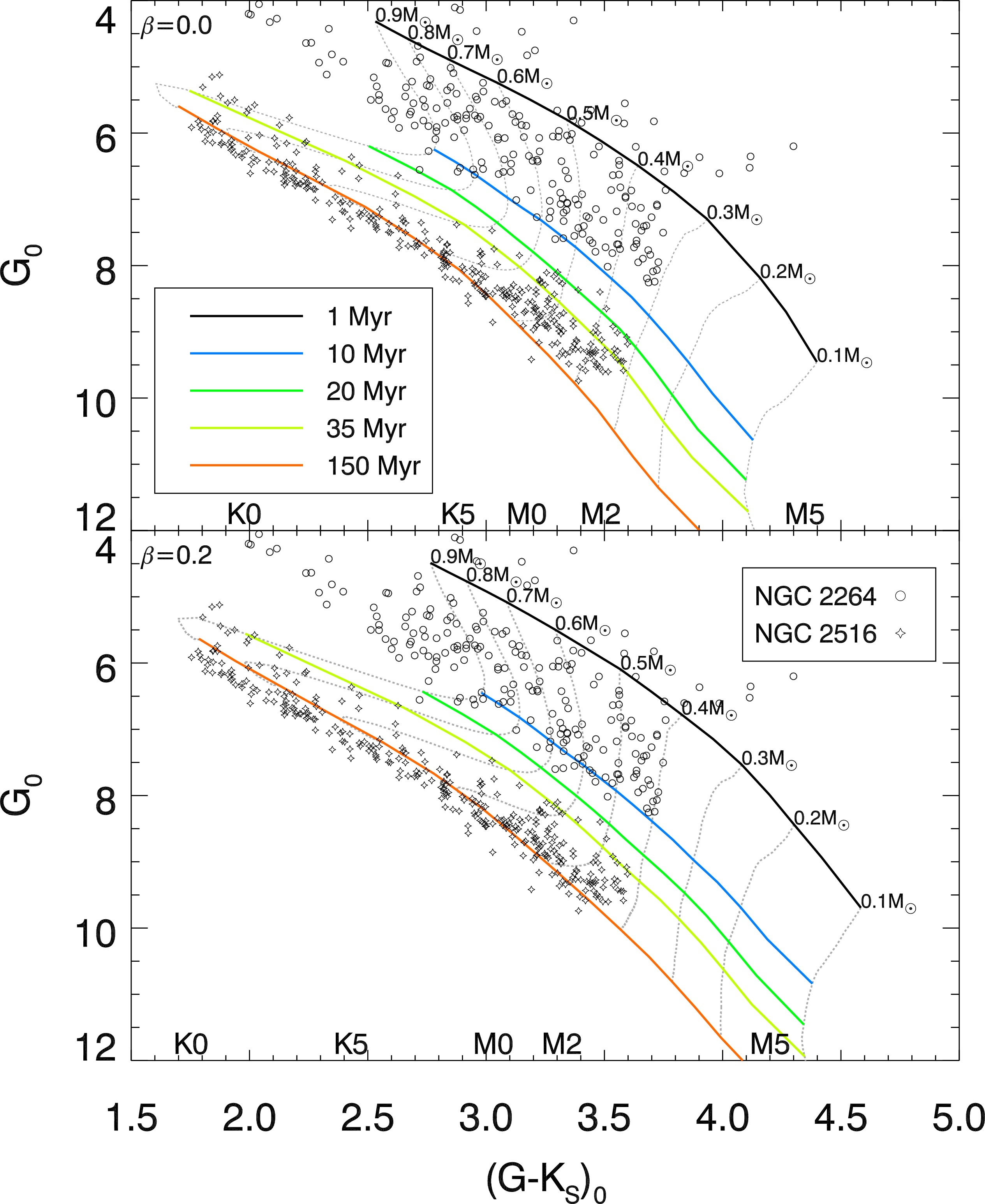}
    \caption{Evolutionary tracks and a set of isochrones between 1 and 150\,Myr plotted on the same CMD axes as Figures~\ref{F_SSM}~and~\ref{F_NSSM} for a set of stellar masses between $0.1 < M/M_{\odot} < 0.8$, using the S20 models with $\beta=0.0$ (top) and $\beta=0.2$ (bottom). Open circles and stars represent NGC~2264 and NGC~2516 members, respectively, whose masses are estimated as $>0.4\,M_{\odot}$~using the S20, $\beta=0.2$~model. The spectral-types on the top and bottom line of text are for the four youngest clusters and for NGC~2516, respectively.}
    \label{F_mass_tracks}
\end{figure}

Along with older ages, adoption of magnetic models would lead to increased inferred masses (Figure~\ref{F_mass_tracks}). The increase is proportionately larger at lower masses. There would then be significant implications for estimating the initial mass function, masses and mass-dependent parameters for low-mass stars and their circumstellar disks, e.g., in NGC~2264 \citep{2012a_Teixeira,2017a_Venuti,2019a_Sousa,2021a_Pearson}, $\lambda$~Ori \citep{2009a_Hernandez,2010a_Hernandez,2012a_Bayo} and NGC~2547 \citep{2004a_Jeffries,2007a_Gorlova}. Improved ages would lead to better-constrained dynamical mass comparisons for individual young systems in nearby MGs, for example, HD~98800 -- a quadruple PMS system recently resolved in the TW Hydrae MG \citep{2021a_Zuniga-Fernandez}.

\subsection{Li and rotation}
\label{S_Discussion_Rotation}

It has been clear for decades that PMS Li depletion does not just depend on the mass and age of a star, but also (to a varying extent) with rotation. The results presented in \S\ref{S_Prot_Analysis} add to the complexity of the current observational picture. The key results are that:

\begin{itemize}
    \item There is a clear correlation between increased Li depletion and slower rotation in the K-type members of the oldest clusters here -- NGC~2547 and NGC~2516. In the case of NGC~2547 this is the first time this correlation has been reported using rotation periods. In the younger clusters there is a weak, but significant correlation between Li depletion and slower rotation in the M-stars of NGC~2264. This correlation is not seen in $\lambda$ Ori (in a smaller sample) and in particular, is not seen in the M-stars of $\gamma$~Vel despite a large dispersion in both Li abundance and rotation period.
    
    \item There is also an accompanying correlation between displacement in absolute magnitude in the CMD and rotation period for the K-type stars of NGC~2547 and NGC~2516 and rotation period, in the sense that faster rotators (or unresolved binary systems) appear to be more luminous. 
    In the younger clusters a similar relationship between rotation and luminosity is seen in the M0-M2 stars, but not in the K-stars.
\end{itemize}

The relative success of the magnetic models in matching both the CMD and the overall Li depletion patterns of these clusters make it tempting to associate any rotation-dependent dispersion of Li and CMD position with differences in magnetic activity associated with the well-known rotation-activity connection in young, low-mass stars. The idea here would be that increased magnetic activity in the most rapidly rotating stars leads to more spotted surfaces or a greater suppression of convection. This in turns leads these stars to be bigger, cooler and have lower central temperatures. This would displace them to higher luminosities and redder colours in the CMD and lead to less Li depletion than equivalent objects with slower rotation and less magnetic activity \citep[e.g.,][]{2015a_Somers}. The rotation-dependent displacements in the CMD are consistent with similar observations of K-stars in the Pleiades and M35, which have a similar age to NGC~2516 \citep{2014a_Kamai, 2016a_Covey, 2021a_Jeffries}. The tripartite correlation between upward movement in the CMD and hence larger radii, reduced Li depletion and rapid rotation is also consistent with what has been found in the Pleiades and M35 \citep{2017a_Somers, 2021a_Jeffries}.

A problem with this model is that most of the Li depletion for K-stars at the ZAMS should take place whilst they are late-K and early M-stars on their Hayashi tracks and thus their younger analogues would be the fully-convective K5-M2 stars of NGC~2264, $\lambda$~Ori and $\gamma$~Vel (see Figure~\ref{F_mass_tracks}). But these objects have rotation periods in the range from $<1$~d to 7~d and would be expected to show fully-saturated levels of magnetic activity throughout the likely epoch of Li depletion \citep[e.g.,][]{2009a_Reiners,2011a_Wright,2020a_Muirhead}. Indeed the levels of X-ray emission seen in NGC~2264, $\lambda$~Ori, $\gamma$~Vel and NGC~2547 are consistent with saturated magnetic activity levels in all K- and M-type members \citep[e.g.,][]{2006b_Jeffries, 2007a_Dahm, 2009a_Jeffries, 2011a_Barrado, 2011a_Franciosini} and for a large fraction of such stars in NGC~2516 \citep{2006a_Pillitteri}. Thus unless magnetic activity, as manifested in a way that changes the structure and heat transport in a PMS star, continues to increase and be rotation dependent beyond the rotation rates at which chromospheric and coronal activity indicators saturate, then it is difficult to see how a clear rotation-dependence is introduced to the Li depletion already at the age of NGC~2547 and even harder to see how it could create any dispersion in Li at the age of NGC~2264. A further problem in interpreting the tripartite correlation is that although inflation by magnetic activity and substantial starspot coverage would shift active stars to higher luminosities and cooler temperatures, so could the presence of an unresolved binary companion. Some part of the correlation between rapid rotation and CMD position could be because the components of binary systems are more likely to be rapidly rotating as has been observed in the Pleiades and Hyades clusters \citep[e.g.,][]{2016a_Stauffer, 2016a_Douglas}.

An alternative interpretation of the observations is that magnetic activity and starspots inhibit Li depletion to approximately the same degree in all young PMS stars, but that additional mixing processes, that are not included in standard models, act to further deplete Li,  and that those mixing processes are more effective in slower rotating stars. Two variants of this have been proposed - one in which the additional mixing accompanies differential rotation associated with early angular momentum loss \citep{2008a_Bouvier} or that convective overshooting into the radiative core becomes important but that rapid rotation prevents the penetration of convective plumes \citep{2017a_Baraffe}.

This interpretation has merit -- it does not require a rotation-activity relation to remain unsaturated at high rotation rates and it explains why the Li-rotation relation is prominent among young (35 Myr) and ZAMS K-stars, which develop radiative cores at ages of 10--20 Myr. A weakness though is that a Li-rotation correlation {\it is} seen here in the M stars of NGC~2264 \citep[and perhaps in the late K-stars,][,though see Appendix~\ref{S_Bouvier16}]{2016a_Bouvier} despite these stars being on fully convective Hayashi tracks (whether considering magnetic or non-magnetic models -- see Figure~\ref{F_mass_tracks}).

The wide dispersion in Li seen in the M-dwarfs of $\gamma$ Vel is difficult to explain in either scenario. These stars should be fully convective and so there is little scope for additional mixing beyond convection and no interface with any radiative core. The differences in Li abundance do not appear to be correlated with rotation at all and in any case, all these stars have saturated levels of magnetic activity. In \S\ref{S_Analysis_gamVel}, the dispersion in Li is accompanied by a dispersion about an isochrone in the CMD that is much larger than can be explained by unresolved binarity. Further, there is a strong correlation between position in the CMD and Li depletion in the sense that the most Li-depleted stars are invariably among the least luminous for their colour. Together, these suggest an explanation for the Li dispersion among the M-stars of $\gamma$~Vel could be that they span a rather broad range of age, perhaps as large as 10~Myr. The possibility of an age spread (or multiple populations) in the $\gamma$~Vel cluster has been discussed before in the context that the isochronal and Li-depletion ages of many of the low-mass stars appear to be significantly older than the eponymous O-star/Wolf-Rayet binary $\gamma^2$~Vel \citep[age $5.5 \pm 1$ Myr,][]{2009a_Eldridge}, which sits at the centre of the cluster \citep[see][]{2009a_Jeffries, 2017a_Jeffries}.

It is worth noting that an age-spread scenario is unlikely to lead to the kind of Li-rotation relation seen in the older clusters and identified in the early M stars of NGC~2264. Firstly, an age spread of $\sim 10$ Myr is not going to make a great deal of difference in populations with average ages $\geq 35$ Myr. Secondly, in a cluster as young as NGC~2264 whose stars are descending Hayashi tracks, then the more Li-depleted and presumably older stars would be smaller, {\it less} luminous and should be {\it faster} rotating as a result of PMS contraction - the opposite of what is seen.

A final possibility, discussed recently by \cite{2021a_Constantino}, is that rapid rotation changes the stability criterion for convection and suppresses mixing in fast-rotating PMS stars. They find that by tuning the amount of convective overshoot, models that incorporate this modified stability criterion are capable of describing why slow-rotating stars are at the bottom of the Li-depletion vs $T_{\rm eff}$ distribution whilst the fastest-rotating stars define the upper envelope. It is unclear how well this mechanism might operate in fully convective stars and so might struggle to explain why a Li dispersion is seen at $<10$ Myr in NGC~2264 and in the M-stars of $\gamma$~Vel, or how this change affects isochronal ages in the CMD. However, like the additional mixing mechanisms discussed above, it has the merit of not relying on a possibly saturated activity-rotation relation to explain the Li-rotation correlation.

\section{Summary}\label{S_Summary}

We have investigated 1,246 high probability, kinematically-selected, K- and M-type members of 5 open clusters at ages from $<10$ to $\sim 125$ Myr (NGC~2264, $\lambda$~Ori, $\gamma$~Vel, NGC~2547 and NGC~2516), that were observed during the GES campaign, covering their evolution from the PMS to the ZAMS. We tested the capability of various evolutionary models, both standard and magnetic, to simultaneously fit both the colour-magnitude diagrams (CMDs) and the equivalent width of the Li~{\sc i}~6708\AA\ feature (EW(Li)) versus colour distributions. The dataset was supplemented with rotation periods estimated from our analysis of TESS lightcurves and/or literature sources, which were used to examine the effects of rotation on CMD position and Li depletion during PMS evolution. The key results are:
\begin{itemize}

    \item Standard model isochrones are unable to simultaneously describe the position of PMS stars in the CMD and the EW(Li) versus colour diagram at the same age. The Li depletion is underestimated at the CMD isochronal ages and occurs at temperatures much cooler than predicted. 

    \item Models incorporating radius inflation caused by surface magnetic fields and/or cool starspots covering 40--60 per cent of the stellar surface provide better fits to the CMD and Li-depletion patterns, but suggest significantly older ages than predicted by standard model isochrones and would predict significantly higher masses for the K- and M-stars of these clusters.
 
    \item We confirm a strong correlation between enhanced Li depletion and slow rotation in the K-type stars of NGC~2516 and find this is also present in NGC~2547 at an age of $\sim 35$ Myr. We also partially confirm the result of \cite{2016a_Bouvier} that there are signs of this correlation in the much younger NGC~2264 cluster, but we find that this is only significant among the M-type stars.
 
   \item The M-dwarfs of the $\gamma$ Vel cluster exhibit a large dispersion in Li abundance but we find no correlation between Li depletion and rotation. This dispersion may be associated with a significant age spread ($\sim 10$ Myr) within the cluster.
   
\end{itemize}

The large discrepancies with the predictions of non-magnetic models suggest that additional input physics is needed in the PMS models and models invoking magnetic activty and starspots are a natural scenario. That the deviations from the CMD isochrone are connected to rotation in the K-stars of NGC~2547 and NGC~2516 and the early-M stars of the younger clusters, in the sense that faster rotators tend to be more luminous, is also broadly in accordance with magnetic models combined with a rotation-activity relationship. It is difficult however to understand why Li depletion should be linked with rotation using these models because all of the low-mass stars in these young clusters either exhibit saturated levels of magnetic activity or would have had saturated levels of magnetic activity whilst Li was being depleted. Alternative models that invoke additional mixing and Li depletion that is more effective in slower rotators are a possibility, but it is difficult to see how these would explain the Li-rotation connection that is observed in the very young, fully-convective PMS stars of NGC~2264 or the wide dispersion of Li depletion in the fully convective M-stars of $\gamma$~Vel. 

\section*{Acknowledgments}

An anonymous referee provided constructive and thoughtful comments which led to a significant improvement in the manuscript. We are grateful to Isabelle Baraffe for supplying a version of the B15 models with finer age resolution. ASB, RDJ and RJJ acknowledge the financial support of the STFC. RB acknowledges financial support from the project PRIN-INAF 2019 ``Spectroscopically Tracing the Disk Dispersal Evolution''. 

Based on data products from observations made with ESO Telescopes at the La Silla Paranal Observatory under programme ID 188.B-3002. These data products have been processed by the Cambridge Astronomy Survey Unit (CASU) at the Institute of Astronomy, University of Cambridge, and by the FLAMES/UVES reduction team at INAF/Osservatorio Astrofisico di Arcetri. These data have been obtained from the Gaia-ESO Survey Data Archive, prepared and hosted by the Wide Field Astronomy Unit, Institute for Astronomy, University of Edinburgh, which is funded by the UK Science and Technology Facilities Council.

This work was partly supported by the European Union FP7 programme through ERC grant number 320360 and by the Leverhulme Trust through grant RPG-2012-541. We acknowledge the support from INAF and Ministero dell' Istruzione, dell' Universit\`a' e della Ricerca (MIUR) in the form of the grant "Premiale VLT 2012". The results presented here benefit from discussions held during the Gaia-ESO workshops and conferences supported by the ESF (European Science Foundation) through the GREAT Research Network Programme.

\section*{Data availability statement}

The data underlying this article are available in the article and in its online supplementary material.

\appendix

\section{Cluster Metallicity}\label{S_Appendix1}

We redetermined the metallicities of the clusters in this paper using the GESiDR5 parameters. The median [Fe/H] and median absolute deviation of [Fe/H] were calculated in each cluster for stars with $P_{\rm 3D}>0.9$ \citep[from][]{2020a_Jackson} and $5000 < T_{\rm eff} {\rm [K]} < 7000$. The values are reported in Table~\ref{T_Metallicities}, where separate results are quoted for targets observed with the UVES and Giraffe spectrographs (note that some targets were observed by both). There does not appear to be any significant difference between the UVES and Giraffe results, so the final values quoted in Table~\ref{T_Clusters} are a combined median and the uncertainties (calculated as the median absolute deviation multiplied by 1.48, assuming a normal distribution, divided by the square root of the total number of targets) represent an internal precision within GES for these homogeneously determined iron abundances. There will be additional external uncertainties to the abundance scale of of 0.05-0.1 dex \citep[see][]{2014a_Smiljanic}.

{\centering
\begin{table*}
\caption{Median iron abundances for the clusters used in this paper. Results are quoted separately for targets observed with the Giraffe and UVES spectrographs. Columns 3 and 6 are the median absolute deviation; columns 4 and 7 are the numbers of stars in each sample.}
\begin{center}
\begin{tabular}{lrrrrrr}
\hline
\hline
Cluster  & \multicolumn{3}{c}{Giraffe} & \multicolumn{3}{c}{UVES}\\
         & Median [Fe/H] & MAD & N     & Median [Fe/H] & MAD & N \\
\hline
     NGC~2264 & $-0.04$ & 0.03 &  18 & $-0.03$ & 0.04 & 12 \\
$\lambda$~Ori & $-0.02$ & 0.04 &   4 & $-0.06$ & 0.03 &  6 \\
 $\gamma$~Vel & $+0.02$ & 0.06 &   8 & $-0.04$ & 0.03 &  4 \\
     NGC~2547 & $+0.01$ & 0.04 &  28 & $-0.01$ & 0.04 & 12 \\
     NGC~2516 & $-0.02$ & 0.05 & 131 & $+0.00$ & 0.03 & 17 \\
    \hline 
\end{tabular}
\end{center}
\label{T_Metallicities}
\end{table*}}

\section{Rotation periods}\label{S_Rotation_Periods}

\subsection{Measuring rotation periods from TESS data}\label{S_TESS}

\begin{figure*}
    \centering
    \includegraphics[width=\textwidth]{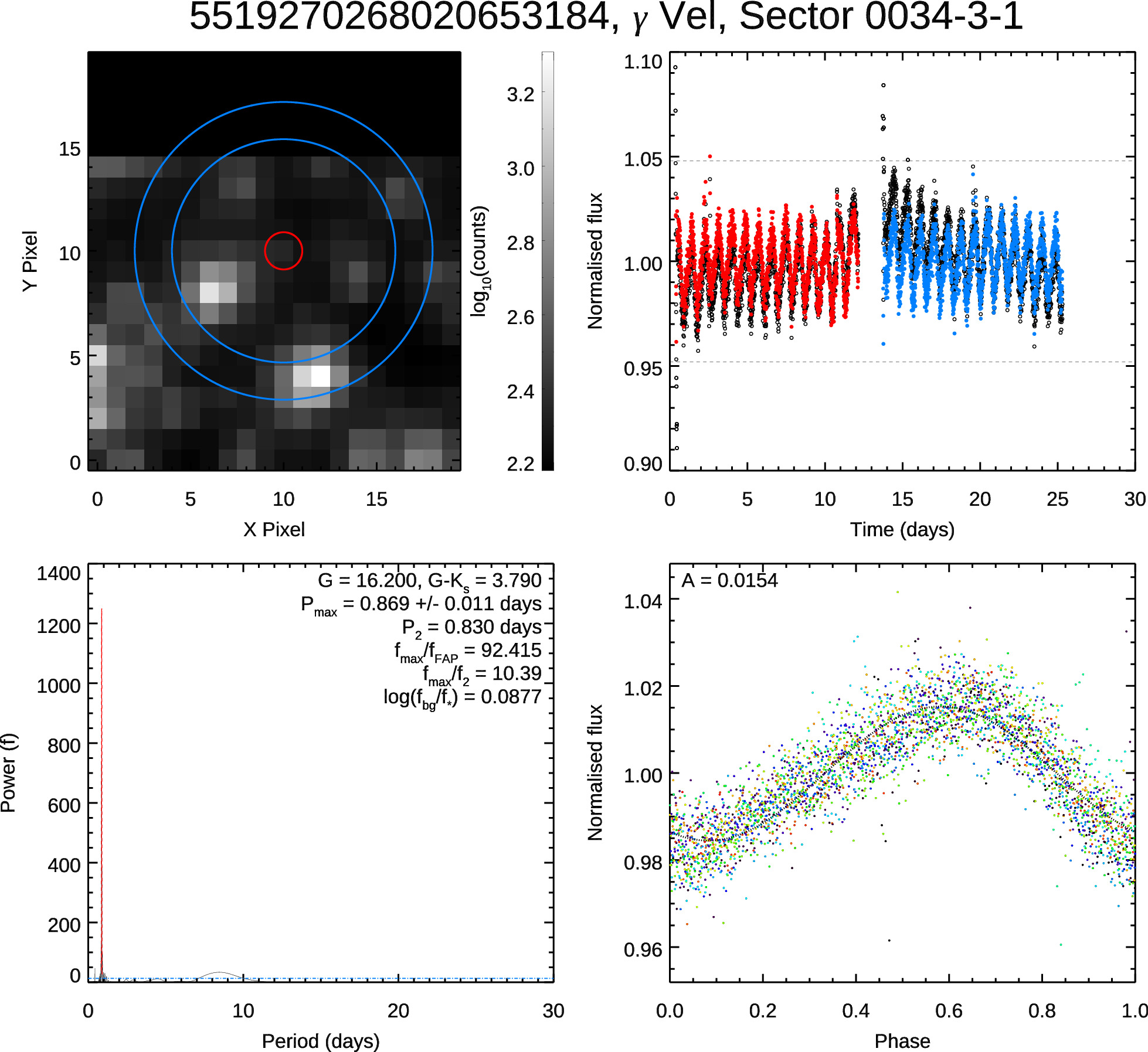}
    \caption{TESS light-curve analysis for a star in our sample with good-quality photometric data. Top-left: TESS 20x20 pixel FFI. The red circle and blue annuli represent the regions where aperture photometry are performed. Top-right: Normalised flux versus time of observation across the duration of the observing sector. Black crosses denote data points that are excluded because they have a TESS quality flag = 1, or have a normalised flux value $>4$ times larger than the median absolute deviation (MAD) from the median flux, denoted by the dashed lines. Black and red/blue dots represent light-curve data before and after applying a second order polynomial detrending algorithm. Bottom-left: Results from the Lomb-Scargle periodogram. The region covered in red represents a Gaussian fit to the maximum power output and the blue horizontal dashed line shows the power threshold of the FAP (at 99 per cent confidence). Bottom-right: The phase diagram, folded on $P_{\rm max}$, where the various colours represent the individual phase cycles. The black dotted line is a fit to the data of the form $y = y_{0} + A\sin{({\omega}t)}$.}
    \label{F_TESS_Good}
\end{figure*}

\begin{figure*}
    \centering
    \includegraphics[width=\textwidth]{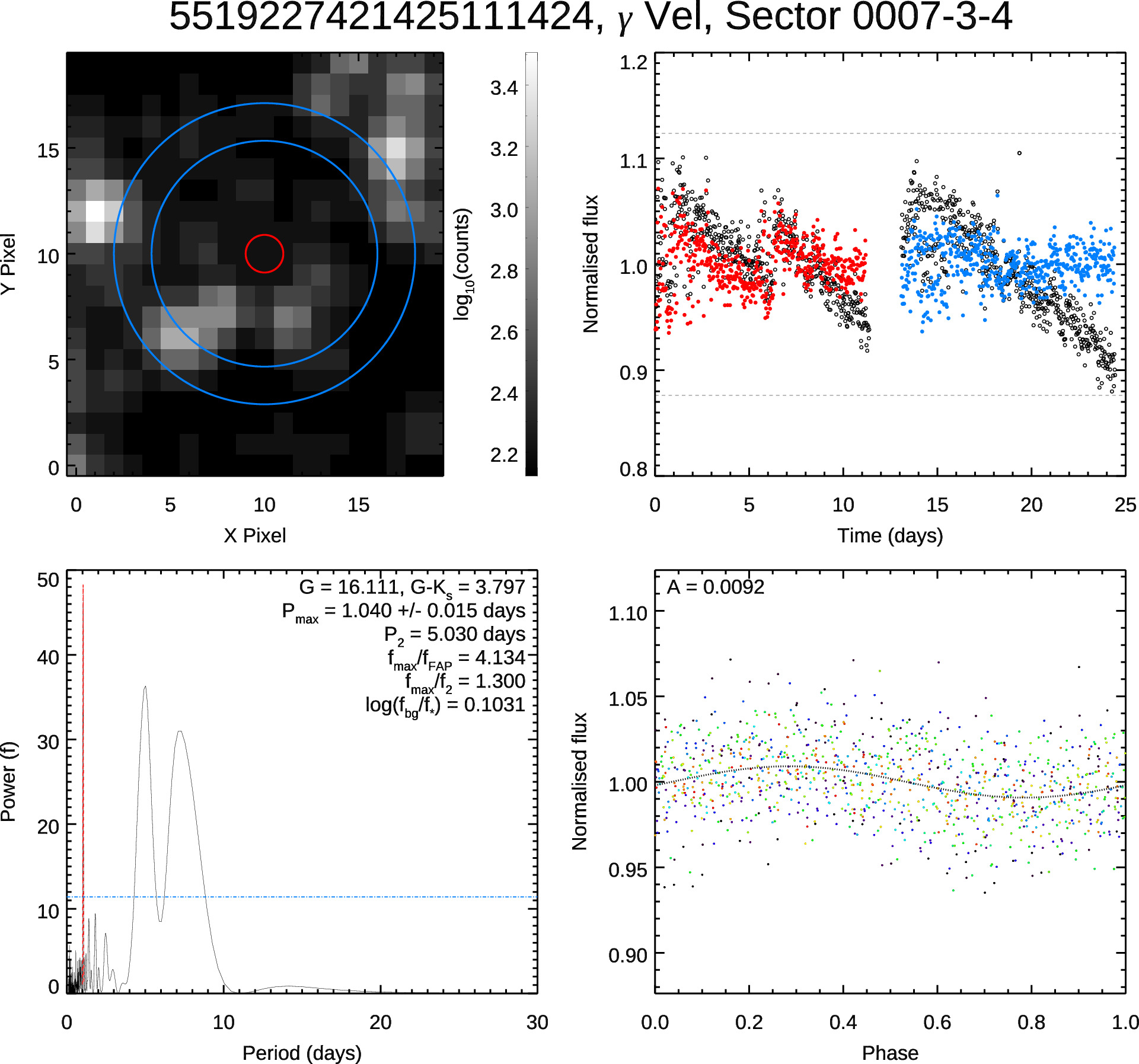}
    \caption{TESS light-curve analysis for a star in our sample with poor-quality photometric data. A description of the panels is given in the caption of Figure~\ref{F_TESS_Good}. The lightcurve (top-right panel) demonstrates where we reject photometric points at the start and end points for the individual components where the flux values are more than 2 MAD from the median flux.}
    \label{F_TESS_Poor}
\end{figure*}

All the selected clusters in Table~\ref{T_Clusters}~were covered in sectors 1--11 during the first year of TESS scientific operations and subsequently at least once more in sectors 27--45. Since $\gamma$~Vel, NGC~2547 and NGC~2516 are located close to the ecliptic south pole, there is often $>1$ sector of TESS data available for objects in these clusters. TESS observes each sector for $\approx$27\,d and full frame images (FFIs) are recorded with a 30\,min cadence and released to the Mikulski Archive for Space Telescopes (MAST). 
For each target we downloaded 20x20 pixel cutouts of the FFIs across all sectors, centered on their GDR2 right ascension ($\alpha$) and declination ($\delta$), using the {\sc tesscut} tool hosted at MAST (\citealt{2019a_Brasseur}\footnote{\url{https://mast.stsci.edu/tesscut/}}) and the recommended {\sc curl} command line procedures. To calculate TESS magnitudes ($m_{\rm TESS}$), aperture photometry was performed on all ($\sim 10^{3}$) individual frames in the image stack produced by {\sc tesscut}, following the procedure described in \cite{2019a_Curtis}. A circular aperture with a radius of 1.0 pixel is used ($\sim$21'' based on the TESS pixel resolution), which is close to the theoretical optimum for flux extraction in the sky-limited domain \citep{1998a_Naylor}, suitable for the relatively faint stars (often in crowded regions) in this work. A sky annulus with inner and outer radii of 6 and 8 pixels is used to fit the (modal) background. Examples of the TESS FFI cutouts are shown in the top-left panel of Figures~\ref{F_TESS_Good}~and~\ref{F_TESS_Poor}.

The lightcurve data ($m_{\rm TESS}$ versus time of observation) were filtered prior to periodogram analysis. Data-points with unexpectedly bright ($m_{\rm TESS} < 0.0$) or faint ($m_{\rm TESS} > 25.0$) photometry or with a quality flag equal to zero were removed. The remaining data points were converted from $m_{\rm TESS}$ to a relative flux and normalised by the median flux value. Data points with normalised flux values ($f_{\rm n}$) that differ from the median $f_{\rm n}$ by more than 4 times the median absolute deviation (MAD) were removed. These filtering steps remove fewer than 1 per cent of the data points for almost every lightcurve used in the analysis.

The remaining data underwent a de-trending process to remove systematics in the lightcurves. Since there are often significant discontinuities in the time-series data, individual fitting coefficients are calculated only for ``data strings'' of at least 20 contiguous points whose neighbouring cadences are less than 10 times the median cadence across the whole sector and where the difference in $f_{\rm n}$ between both the start and end point of the string and the median $f_{\rm n}$ across the entire sector is less than 2 MAD. The fitting type, either linear or parabolic, is determined by the function that provides the best least-squares fit to the string. Example TESS lightcurves are shown in the top-right panel of Figures~\ref{F_TESS_Good}~and~\ref{F_TESS_Poor}, where the individual data strings are coloured alternately in red and blue.

A Lomb-Scargle periodogram analysis was performed on the processed lightcurves, sampled between 0.1 and 30 days. A sub-sample of 50 objects were selected to measure the power values corresponding to false-alarm probabilities (FAPs) at the 99 per cent confidence interval ($f_{\rm FAP}$) using two methods: (a) equation 18 in \cite{1982a_Scargle}, where the number of independent frequencies is calculated using equation 3 in \cite{1986a_Horne} and (b) 1500 white noise simulations. The median difference between $f_{\rm FAP}$ values calculated from methods (a) and (b) is $-0.985$ (MAD = $0.098$), which is very small compared with power output values from the peak of the power spectrum ($f_{\rm max}$), which are typically several hundreds (see, e.g., Figure~\ref{F_TESS_Good}). To avoid large computational expense in performing white noise simulations for thousands of lightcurves, we calculated $f_{\rm FAP}$ using method (a) and subtracted 0.985 from the value. 

A Gaussian fit was centered around $f_{\rm max}$, where the period derived from each lightcurve, $P_{\rm TESS}$, was calculated as the centroid of the peak, and the measurement uncertainty ($\sigma_{P_{\rm TESS}}$) as the full width half maximum. The value corresponding to the second largest power spectrum maximum ($f_{2}$) and its corresponding period ($P_{2}$) were also calculated. The background contamination (see $\S$\ref{S_TESS_Background}) and the ratios $f_{\rm max}/f_{\rm 2}$ and $f_{\rm max}/f_{\rm FAP}$ ($\S$\ref{S_TESS_Comparison}), were used as metrics for choosing objects with reliable $P_{\rm TESS}$ values. Examples of periodograms are shown in the bottom-left panels of Figures~\ref{F_TESS_Good}~and~\ref{F_TESS_Poor}.

The lightcurve data were phase-folded using $P_{\rm TESS}$ and fitted with a sine function. An example of a TESS lightcurve is shown in the top-right panel of Figures~\ref{F_TESS_Good}~and~\ref{F_TESS_Poor}. This procedure allowed us to visually check for features in the lightcurve that may not be related to the periodicity.

For all lightcurves with reliable data, the outputs from the TESS analysis, namely $P_{\rm TESS}$, the TESS sector, $f_{\rm max}/f_{\rm FAP}$, $f_{\rm max}/f_{2}$ and $\log{(f_{\rm bg}/f_{*})}$ are presented in Table~\ref{T_TESS}. The final rotation period from the TESS data, $P_{\rm TF}$, is simply $P_{\rm TESS}$ in the case where just one sector is available. The algorithm to calculate $P_{\rm TF}$ for targets with 2 or more sectors is as follows: if the standard deviation of the $P_{\rm TESS}$ values is less than the mean value of $\sigma_{P_{\rm TESS}}$, $P_{\rm TF}$ is the error-weighted mean, otherwise the lightcurve whose $P_{\rm TESS}$ value is furthest the from the mean value is removed and the process is reiterated until either the first condition is satisfied or if there are only 2 $P_{\rm TESS}$ values remaining (or only 2 to begin with) and the condition is still not satisfied, then the $P_{\rm TESS}$ with the largest corresponding $f_{\rm max}/f_{2}$ is chosen as $P_{\rm TF}$. Of the 296 targets with 2 or more $P_{\rm TESS}$ measurements, 257 have 2 or more periods that are consistent, indicating that 87 per cent of the TESS lightcurves provide consistent $P_{\rm TESS}$ values.

\subsection{TESS Background contamination}\label{S_TESS_Background}

In the TESS images, a square enclosing half the energy from a point source has a side of length 21 arcsec (1 pixel), therefore source contamination and background regions defined in $\S$\ref{S_TESS} that may contain bright objects in crowded cluster fields may lead to an unreliable estimate of $P_{\rm TESS}$.

To quantify the ratio of potential contaminating flux within the source aperture, compared to the source data, we queried GDR2 for all objects within a 5 pixel radius of the target. From equation 3b-10 in \cite{1965a_Biser}, the contribution from these sources to the circular target aperture is (assuming a Gaussian point spread function [PSF]):
\begin{equation}\label{E_1}
f_{\rm bg} = \sum_{i} 10^{-0.4 G_{i}}\, e^{-t} \sum_{n=0}^{n\to{\infty}}{\Bigg\{\frac{t^{n}}{n!}\bigg[1-e^{-s}\sum_{k=0}^{n}{\frac{s^{k}}{k!}}}\bigg]\Bigg\}\, ,
\end{equation}
where the outer summation is over all potential contaminants, $G$ is the GDR2 $G$-band magnitude, $s =R^{2}/2\sigma^{2}$ and $t= D^{2}/2\sigma^{2}$, $R$ is the aperture radius, $D$ is the distance between the contaminating object and the target aperture centre and $\sigma$ is the FWHM of the TESS PSF, which is approximately 0.65 pixels. This can be compared with the flux from the target object within $R$, which is approximated by a Rayleigh distribution:
\begin{equation}\label{E_2}
f_{\rm *} = 10^{-0.4\,G}\,\left(1 - e^{-s}\right).
\end{equation}
Most of the potential contaminants are fainter than the target and are unlikely to compromise the lightcurve. Individual contaminants that contribute similar, or larger flux may lead to a noisier lightcurve and/or an unreliable period for the target star.

A periodogram analysis is made for each potential contaminant that contributes $>$10 per cent of source flux, centered on their positions, using the same sky background as for the target star. If this yields a period that is within one error bar of the original target period then the measured $P_{\rm TESS}$ may be that of the contaminant. If the contaminant contributes more flux than the target and is unresolved from the target (i.e. within 1 pixel) then $P_{\rm TESS}$ is flagged as unreliable. If the separation is greater than 1 pixel then $P_{\rm TESS}$ is still flagged as unreliable if the amplitude of the phase-folded lightcurve is more than half that of the original target lightcurve. A periodogram analysis was also carried out on the flux in the background annulus, but in no cases did this lead to a period similar to that found in a target. From 6787 lightcurves (prior to further tests described in $\S$\ref{S_TESS_Comparison}), 1499 were discarded as potentially unreliable due to contamination. A gallery of the 5288 lightcurves that satisfy these contamination criteria are provided as an electronic supplement in the same format as Figures~\ref{F_TESS_Good}~and~\ref{F_TESS_Poor}.

{\centering
\begin{table}
\begin{center}
\begin{tabular}{p{0.2cm}p{1.7cm}p{0.8cm}p{0.6cm}p{0.5cm}p{0.6cm}p{0.5cm}}
\hline
\hline
$\#$                     & $P_{\rm TESS}$    & Sector & $C_{1}$ & $C_{2}$ & $C_{3}$ & ticks \\
 & (days)            &        &                           &                     &   & \\
\hline
1  & $2.455 \pm 0.090$ & 6-1-3  & $14.78$ & $2.15$  & $+0.24$ & \cmark\cmark\cmark \\
   & $2.465 \pm 0.093$ & 33-1-3 & $32.48$ & $1.29$  & $+0.24$ & \cmark\xmark\cmark \\
2  & $5.766 \pm 0.567$ & 6-1-3  & $12.58$ & $1.76$  & $-0.56$ & \cmark\xmark\cmark \\
   & $5.577 \pm 0.431$ & 33-1-3 & $49.13$ & $2.76$  & $-0.56$ & \cmark\cmark\cmark \\
3  & $7.152 \pm 1.522$ & 6-1-3  & $6.40$  & $1.75$  & $+1.73$ & \cmark\xmark\xmark \\
4  &                   &        &         &         &         &                    \\
5  & $5.586 \pm 0.539$ & 6-1-3  & $2.75$  & $1.57$  & $-0.14$ & \xmark\xmark\cmark \\
   & $8.672 \pm 1.148$ & 33-1-3 & $63.66$ & $3.16$  & $-0.14$ & \cmark\cmark\cmark \\
6  &                   &        &         &         &         &                    \\
7  & $3.858 \pm 0.227$ & 6-1-3  & $3.00$  & $2.46$  & $-0.42$ & \xmark\cmark\cmark \\
   & $0.111 \pm 0.001$ & 33-1-3 & $0.83$  & $1.69$  & $-0.42$ & \xmark\xmark\cmark \\
8  & $3.734 \pm 0.214$ & 6-1-3  & $16.71$ & $3.64$  & $-0.19$ & \cmark\cmark\cmark \\
   & $3.687 \pm 0.193$ & 33-1-3 & $5.26$  & $1.01$  & $-0.19$ & \cmark\xmark\cmark \\
9  &                   &        &         &         &         &                    \\
10 & $0.858 \pm 0.012$ & 6-1-3  & $35.80$ & $11.07$ & $-0.37$ & \cmark\cmark\cmark \\
   & $0.858 \pm 0.010$ & 33-1-3 & $92.65$ & $6.60$  & $-0.37$ & \cmark\cmark\cmark \\
\hline
\end{tabular}
\caption{Period data for all TESS lightcurves that are not flagged as potential contaminants in $\S$\ref{S_TESS_Background}. Column 1 is the designated number corresponding to the source identifier in Table~\ref{T_Data}. Columns 2--6 contain data from the TESS analysis in this work -- the calculated rotation period, the TESS sector, camera and CCD and the three threshold criteria described in $\S$\ref{S_TESS_Comparison}, where $C_{1} = f_{\rm max}/f_{\rm FAP}$, $C_{2} = f_{\rm max}/f_{2}$ and $C_{3} = \log (f_{\rm bg}/f_{*})$. Column 7 indicates whether the criteria $C_{1}$, $C_{2}$ and $C_{3}$ pass (\cmark) or fail (\xmark), respectively. Only the first 10 sources are listed here. The entire table is available in electronic format.}
\label{T_TESS}
\end{center}
\end{table}}

\subsection{Comparison of TESS rotation periods with other sources}\label{S_TESS_Comparison}

To further test the reliability of our TESS-derived rotation periods and as a source of additional data, we compiled literature rotation periods for objects in our parent sample. We find rotation periods available for 164, 49 and 184 objects in NGC~2264 (from the compilations of \citealt{2014a_Cody}~and~\citealt{2016a_Bouvier}), NGC~2547 \citep{2008a_Irwin} and NGC~2516 (\citealt{2007a_Irwin}, I07 and~\citealt{2020a_Fritzewski}, F20), respectively. Only the data from F20 contain measurement uncertainties. There are 25 NGC~2516 objects with a rotation period in both I07 and F20. For these objects the final selected rotation period is either the mean of the two measurements if they are in agreement (20 objects), or from F20 if the periods do not match (5 objects).

Even after removing potential unreliable periods due to contamination ($\S$\ref{S_TESS_Background}) there may be further unreliable TESS lightcurves. This is because of (i) low signal-to-noise for fainter targets, (ii) the limited duration of the TESS observations compared with the longer rotation periods in the sample, (iii) contamination of the TESS lightcurves by surrounding objects. To characterise these problems, three metrics were defined: the ratios $f_{\rm max}/f_{\rm 2}$, $f_{\rm max}/f_{\rm FAP}$ and $\log({f_{\rm *}/f_{\rm bg}})$ (see $\S$\ref{S_TESS}).

A grid of threshold values for these ratios was defined and at each grid point a comparison was made between $P_{\rm TF}$ (where all the individual $P_{\rm TESS}$ values satisfy the background contamination criteria and these three metrics) and all objects with a corresponding rotation period in the literature ($P_{\rm lit}$). Satisfactory agreement is defined as $|P_{\rm TF}-P_{\rm lit}|/\sigma_{P_{\rm TF}} < 3.0$ and $|P_{\rm TF}-P_{\rm lit}| < 2.0\,$days. The optimal set of thresholds are those that provide a high fraction of matching periods whilst not rejecting a significant amount of data that would prove useful in the analysis. 

After visual assessment of all the thresholds in the grid, we settled on $f_{\rm max}/f_{\rm 2} > 2.0$, $f_{\rm max}/f_{\rm FAP} > 5.0$ and $\log({f_{\rm bg}/f_{\rm *}}) < 0.4$, since this allows enough targets to be analysed from a statistical point of view (for the analysis in $\S$\ref{S_Prot_Analysis}) whilst recovering a reasonable fraction of matching periods with literature values. Whilst we considered a receiver operator characteristic (ROC) curve analysis to select the optimal thresholds, ultimately we decided to choose them manually because (a) the $P_{\rm lit}$ values are not necessary all equal to the true rotation period and (b) there are multiple criteria that determine the number of targets and recovery rate. From the initial sample of 1109 objects that passed the background contamination tests in $\S$\ref{S_TESS_Background}, 477 passed these thresholds for at least one sector of data and of those with a published $P_{\rm lit}$ (222 objects), 73 per cent were in agreement. A plot comparing $P_{\rm TF}$ and $P_{\rm lit}$ values is presented in Figure~\ref{F_Period_Comparison}. Greater levels of agreement were possible by increasing the thresholds, but at the expense of discarding a much larger fraction of the TESS data. 

\begin{figure}
    \centering
    \includegraphics[width=0.45\textwidth]{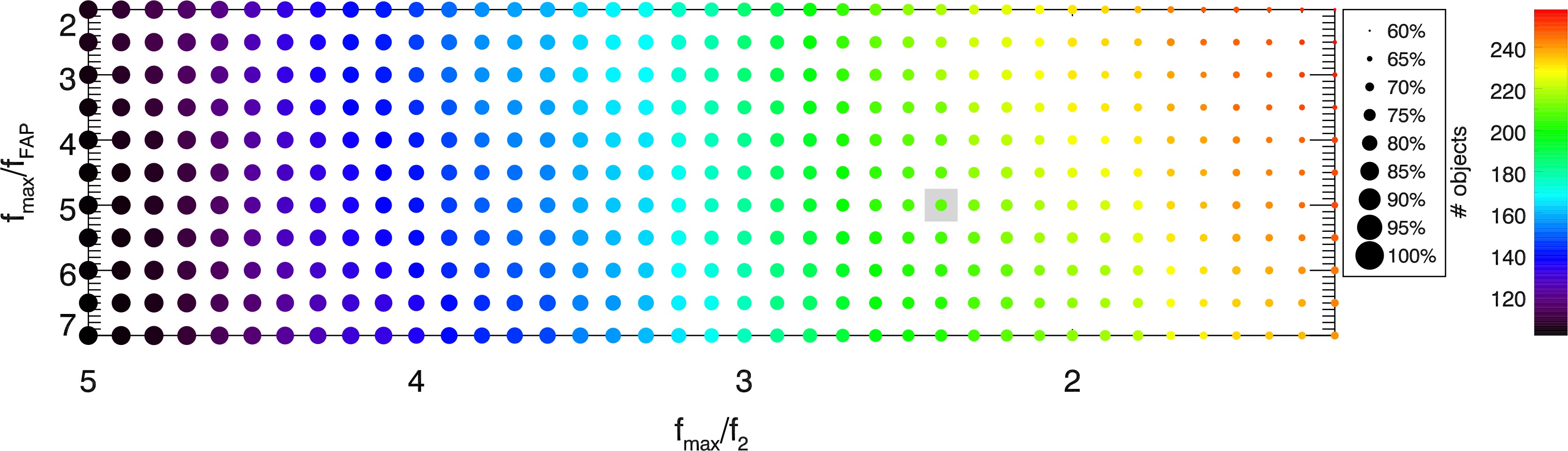}
    \caption{The number of objects with periods in both TESS and the literature (represented by symbol size) and the percentage with matched values (using the criteria in $\S$\ref{S_TESS_Comparison}, represented on a colour scale) fixed at $f_{\rm max}/f_{\rm FAP} = 5.0$ for a range of threshold values for $f_{\rm max}/f_{2}$ and $\log(f_{\rm bg}/f_{\rm *})$. The location of the grey square represents the criteria we choose for our analysis.}
    \label{F_TESS_Thresholds}
\end{figure}

It should be noted that perfect agreement is not expected. The rotation periods for NGC~2547 and NGC~2516 are exclusively from ground-based photometric surveys, and it is possible that some of these data are less reliable as they were not observed over a long baseline and/or may be subject to $\sim 1$ day aliasing effects. The NGC~2264 data are mainly from space-based observations with CoRoT \citep{2007a_Cieza,2013a_Affer,2017a_Venuti}, though some measurements have their origins in ground-based data \citep{2004a_Lamm,2004a_Makidon}. For NGC~2264 objects with both space- and ground-based literature periods, there is only 80 per cent agreement between these. It is always possible that the measured period is a harmonic of the principle frequency caused by a particular distribution of brightness inhomogeneities on the stellar surface. 

\begin{figure}
    \centering
    \includegraphics[width=0.45\textwidth]{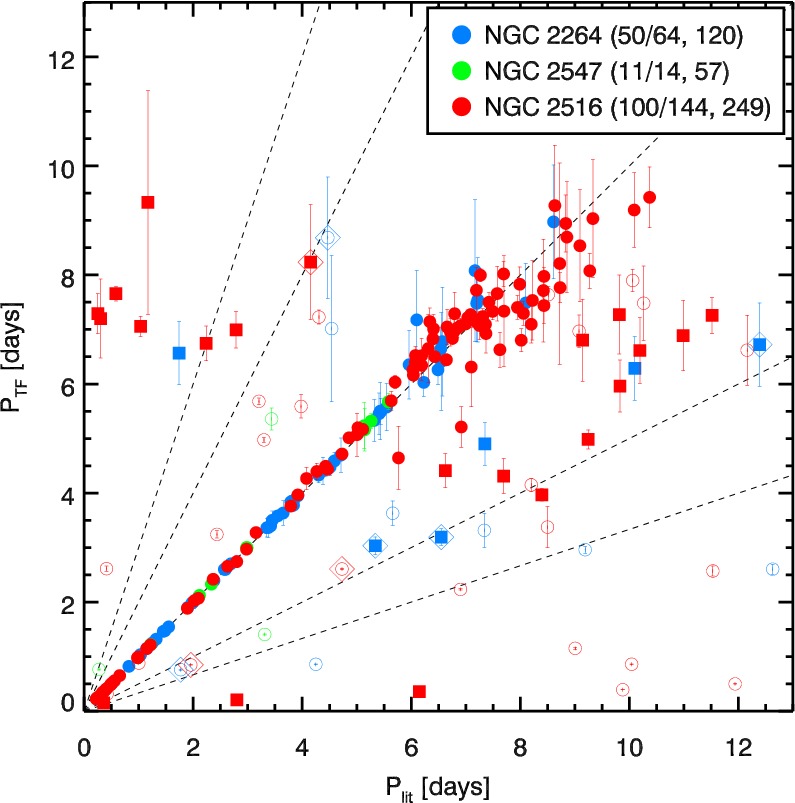}
    \caption{Rotation period derived from TESS analysis ($P_{\rm TF}$) versus the rotation period from the literature source ($P_{\rm lit}$) for targets that satisfy the threshold values adopted in our analysis ($f_{\rm max}/f_{\rm 2} > 2.0$, $f_{\rm max}/f_{\rm FAP} > 5.0$ and $\log({f_{\rm bg}/f_{\rm *}}) < 0.4$). Filled circles represent objects that satisfy the matching criteria described in $\S$\ref{S_TESS_Comparison}. Open circles and squares are targets that fail the matching criteria where the adopted period is from TESS and the literature, respectively. The targets marked with large diamonds have $P_{\rm lit} \simeq 2\,P_{\rm TF}$, where $P_{\rm lit}$ is the adopted period. The dotted lines represent gradients of 1/3, 1/2, 1, 2 and 3. The fractional values in the legend represent the number of targets that satisfy matching criteria (in the numerator) and the total number of targets available for testing (in the denominator) and the integer value represents the total number of targets with $P_{\rm TESS}$ and $P_{\rm lit}$ available without threshold values applied.}
    \label{F_Period_Comparison}
\end{figure}

\subsection{Final period determination, $P_{\rm rot}$}\label{S_Final_Period_Determination}
The final adopted rotation period, to be used in all subsequent analysis in this work, is denoted as $P_{\rm rot}$. Where we have only (a reliable) $P_{\rm TF}$ or a $P_{\rm lit}$, then that is adopted as $P_{\rm rot}$. In 61 cases, $P_{\rm TF}$ and $P_{\rm lit}$ disagree (see Figure~\ref{F_Period_Comparison}): 14 in NGC~2264, 3 in NGC~2547 and 44 in NGC~2516. For these targets the lightcurves were visually inspected. For 27 targets the TESS lightcurves are noisy and/or low-amplitude and we consider $P_{\rm TF}$ less reliable, therefore $P_{\rm lit}$ values were used. The remaining 34 show clear periodic signals (for all but one of these targets there are consistent period measurements from multiple sectors), of which 13 have at least one lightcurve with $P_{2} \sim P_{\rm lit}$. Four of these 34 targets have $P_{\rm lit} \simeq 2P_{\rm TF}$ (and also 4 of the 27 targets with noisy and/or low-amplitude lightcurves) where it is likely that TESS has mistakenly identified half the true period because of the way spots are distributed on the surfaces, therefore $P_{\rm lit}$ was adopted. For the remaining 28 we assume that $P_{\rm TF}$ is the correct value.

Values of $P_{\rm TF}$, $P_{\rm lit}$ and $P_{\rm rot}$ are given in Table~\ref{T_Periods}, as well as a reference code which describes the selection process used to choose $P_{\rm rot}$. We note that many of the GES targets and most in $\gamma$~Vel, NGC~2547 and NGC~2516 have 2 or more sectors of TESS data with consistent $P_{\rm rot}$ values, improving their reliability. In total, we obtain rotation periods for 698 (237) sources, of which 223 (74), 65 (65), 46 (46), 73 (11) and 291 (41) belong to NGC~2264, $\lambda$~Ori, $\gamma~$Vel, NGC~2547 and NGC~2516, respectively, where the values in parentheses are the number of sources where only a $P_{\rm TF}$ value is available (i.e., a period recorded for the first time). In total, there are 425 sources where the $P_{\rm rot}$ value is from $P_{\rm TF}$.

From the 39 targets with 2 discrepant TESS periods, 21 have a $P_{\rm lit}$~measurement, 10 of which are in agreement with our selected $P_{\rm TF}$. Finally, from the 181 objects with only one TESS lightcurve, 52 have $P_{\rm lit}$~values, where 33 are in agreement. Since neither sources of $P_{\rm rot}$~provide the ``ground truth'', we choose not to alter our $P_{\rm rot}$ determination.

{\centering
\begin{table}
\begin{center}
\begin{tabular}{llllll}
\hline
\hline
$\#$ & $P_{\rm TF}$    & $N$ & $P_{\rm lit}$ & $P_{\rm rot}$     & $C$  \\
     & (days)            &              & (days)        & (days)            &       \\
\hline
1  & $2.455 \pm 0.090$ & 1 &              & $2.455 \pm 0.090$ & 1 \\
2  & $5.577 \pm 0.431$ & 1 & $5.55$, V+16 & $5.577 \pm 0.431$ & 3 \\
3  &                   & 0 &              &                   & 0 \\
4  &                   & 0 &              &                   & 0 \\
5  & $8.672 \pm 1.148$ & 1 &              & $8.672 \pm 1.148$ & 1 \\
6  &                   & 0 &              &                   & 0 \\
7  &                   & 0 &              &                   & 0 \\
8  & $3.734 \pm 0.214$ & 1 &              & $3.734 \pm 0.214$ & 1 \\
9  & $6.875 \pm 0.594$ & 1 &              & $6.875 \pm 0.594$ & 1 \\
10 & $0.858 \pm 0.005$ & 2 & $4.25$, V+16 & $0.858 \pm 0.005$ & 5 \\

\hline
\end{tabular}
\caption{TESS and literature rotation periods and the final adopted rotation period to be used for analysis. Column 1 is the source identifier from Gaia DR2. Columns 2 and 3 provide the rotation period derived in the TESS analysis ($P_{\rm TF}$) and the number of lightcurves, $N$, that are used for calculating $P_{\rm TF}$. Column 4 provides the literature rotation period ($P_{\rm lit}$) and the source reference: A+13 = \protect\cite{2013a_Affer}; F+20 = \protect\cite{2020a_Fritzewski}; I+07 = \protect\cite{2007a_Irwin}; I+08 = \protect\cite{2008a_Irwin}; L+04 = \protect\cite{2004a_Lamm}; M+04 = \protect\cite{2004a_Makidon}; V+17 = \protect\cite{2017a_Venuti}. Column 5 is the final rotation period ($P_{\rm rot}$) to be used for analysis and column 6 denotes a code number describing how the final period was chosen: (0) no period data available (neither $P_{\rm TF}$ nor $P_{\rm lit}$); (1) one lightcurve from TESS (no $P_{\rm lit}$ available); (2) a $P_{\rm lit}$ value available (no $P_{\rm TF}$); (3) both $P_{\rm TF}$ and $P_{\rm lit}$ are in agreement, $P_{\rm rot} = P_{\rm TF}$; (4) $P_{\rm TF}$ and $P_{\rm lit}$ are in disagreement, $P_{\rm rot} = P_{\rm lit}$; (5) $P_{\rm TF}$ and $P_{\rm lit}$ are in disagreement, $P_{\rm rot} = P_{\rm TF}$; (6) $P_{\rm TF} \sim 2P_{\rm lit}, P_{\rm rot} = P_{\rm lit}$; (7) two or more lightcurves from $P_{\rm TESS}$ and no $P_{\rm lit}$ is available. Only the first 10 sources are listed here. The entire table is available in electronic format.}
\label{T_Periods}
\end{center}
\end{table}}

\subsection{How reliable are the $P_{\rm rot}$ values?}\label{S_Prot_Reliable}
Whilst the majority of stars with 2 or more $P_{\rm rot}$ measurements (either from multiple TESS lightcurves or comparative literature sources) are in good agreement, a percentage of these are at odds. Figure~\ref{F_Period_Comparison}~shows numerous examples where TESS measures a long period ($\gtrsim7$ days), but the literature value $P_{\rm rot}$ is much shorter ($\lesssim2$ days), or vice versa. Our aim here is to estimate the fraction of stars that could be considered as having unreliable $P_{\rm rot}$ values because of clear differences in the measurements. We split these into three categories:

\begin{enumerate}
    \item Targets with only literature values available (corresponding to $C=2$ in Table~\ref{T_Periods}), where there are at least 2 independent measurements that are in disagreement (3 out of 27 cases).
    \item Targets for which only TESS data is available and there are at least 2 lightcurves. We count potentially unreliable $P_{\rm rot}$ values as those where the algorithm to calculate $P_{\rm TF}$ (see last paragraph in $\S$\ref{S_TESS}) has to use 2 discrepant $P_{\rm TESS}$ measurements (18 out of 126 cases).
    \item Targets where TESS and literature periods disagree (i.e., the mismatches on Figure~\ref{F_Period_Comparison}), excluding 23 which have $\geq2$ TESS periods that are in agreement with each other (but not the literature) and exhibit no indications of unreliability on visual assessment of the lightcurves (38 out of 218 cases).
\end{enumerate}

This gives our estimated fraction of potentially unreliable $P_{\rm rot}$ values as 59/371, approximately 16 per cent. However, this fraction is not equivalent to the number of targets that have incorrect periods. We expect the true number of incorrect $P_{\rm rot}$ values to be lower because the decisions made in the final selection of $P_{\rm rot}$ are often weighted based on features of the lightcurve data (e.g., $f_{\rm max}/f_{\rm 2}$).

\section{Comparing EW(Li) and rotation trends in NGC~2264 with Bouvier et al. 2016}\label{S_Bouvier16}

Using GES Li measurements from its 4th internal data release (GESiDR4), \citet[][herein B16]{2016a_Bouvier} reported a weak, negative correlation between Li-depletion and rotation in NGC~2264. The authors calculated a least-squares linear fit of EW(Li) versus $T_{\rm eff}$ for targets with $3800 < T_{\rm eff} < 4400$\,K, that are listed in the ``CSI 2264" database \cite{2014a_Cody}. The methodolology adopted by B16 of comparing residual values of EW(Li) with $P_{\rm rot}$ (see their figure 4) is very similar to the approach in $\S$\ref{S_Analysis_EWLi}. Figure~\ref{F_Prot_EWdiffN}~indicates no significant correlation between $\delta$EW(Li) and $P_{\rm rot}$ for K5-M0 stars but a significant, weak negative correlation amongst the M-stars. In order to provide a more direct comparision with B16 we selected the targets chosen by B16 and also those occupying a similar range of $(G-K_{\rm s})_0$, and we defined two samples as follows:

\begin{enumerate}
\item For the ``B16 sample'' we collected (dereddened, using the $E(B-V)$ in Table~\ref{T_Clusters}) $G$ and $K_{\rm s}$ photometry for the 63 objects with $3800 < T_{\rm eff} < 4400\,$K in table 1 of B16 (and not rejected from their EW(Li) fitting process), and calculate $\delta$EW(Li) using a least-squares fit between the EW(Li) given in B16 and $(G-K_{\rm s})_0$. The $P_{\rm rot}$ values are those from from B16.
\item The sample based on ``this work'' were selected using the  131 NGC~2264 targets in our membership list (with $P_{\rm 3D} > 0.97$) that have a $P_{\rm rot}$ measurement (in Table~\ref{T_Periods}), are not classified as strong accretors, and lie within 2 scaled median absolute deviation values of the median $G-K_{\rm s}$ of the sample defined in step 1 ($2.75 < (G-K_{\rm s})_0 < 3.61$, corresponding roughly to spectral types of K4-M2). The $\delta$EW(Li) and $P_{\rm rot}$ values for this sample were taken from this work.
\end{enumerate}

Of the 63 targets considered by B16, only 50 are present in our list of NGC~2264 members. The 13 missing targets have $P_{\rm 3D}<0.97$, below the threshold for membership criteria in this work. Most of these have proper motions consistent with membership, but with discrepant RVs, suggesting they are either real members comprised of RV-variable, close multiple systems, or non-members whose proper-motion vectors are coincidentally aligned with NGC~2264. There are however more stars in the same approximate $T_{\rm eff}$ range in this paper because of the addition of many more rotation periods and a more complete analysis of the GES Li data.

\begin{figure}
    \centering
    \includegraphics[width=0.45\textwidth]{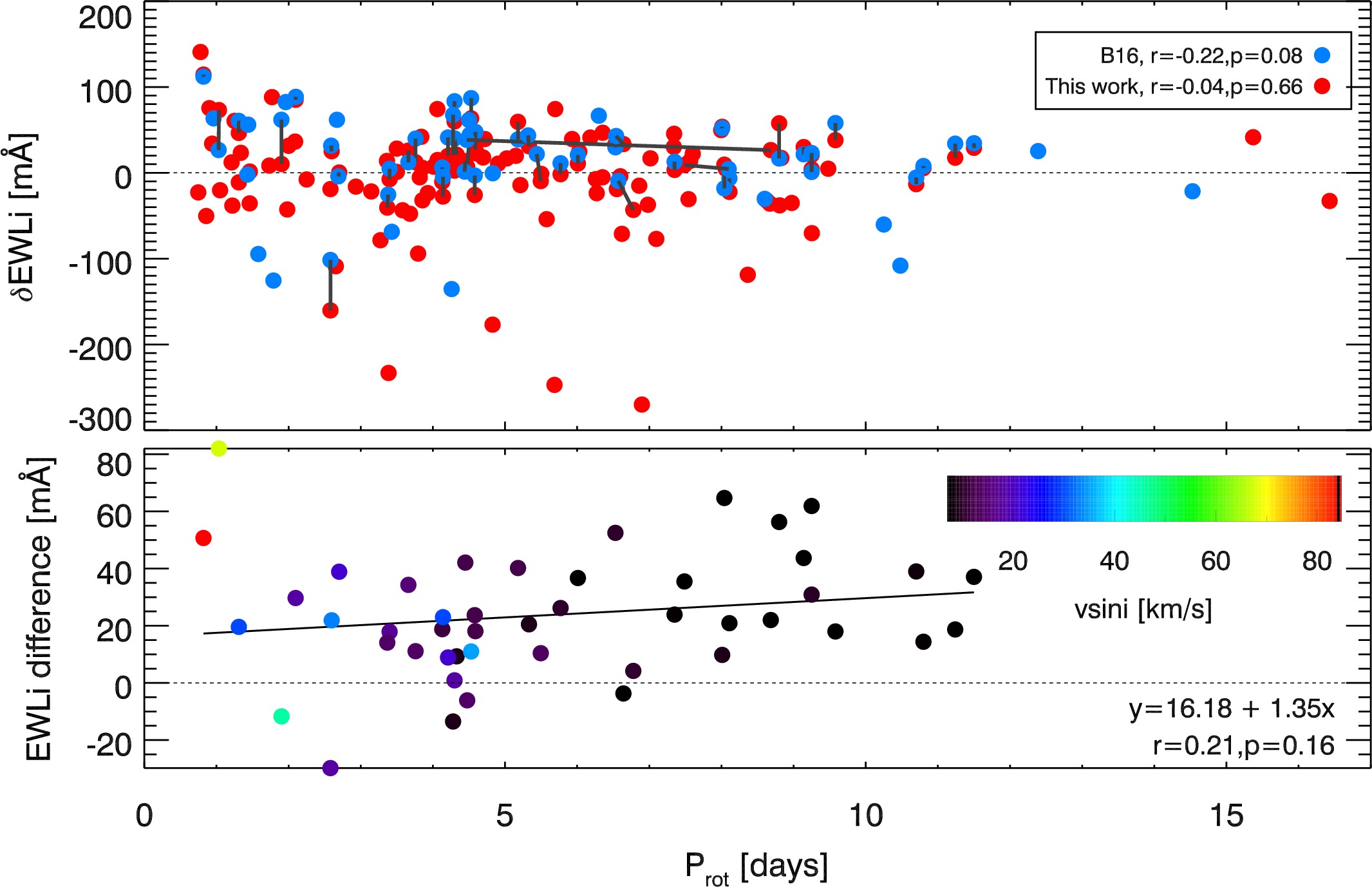}
    \caption{Top: EW(Li) offsets ($\delta$EW(Li)) versus $P_{\rm rot}$ where blue filled circles represent targets in B16 that have been converted into $\delta$EW(Li) by subtracting a mean relationship between EW(Li) and $(G-K_{\rm s})_0$. Red filled circles are NGC~2264 objects in our sample within 2 scaled median absolute deviation (MAD) values of the median $(G-K_{\rm s})_0$ value of the B16 sample, calculated from the linear fitting procedure described in $\S$\ref{S_Prot_Li}. Grey lines connect targets common to both samples. Bottom: The difference in EW(Li) measured in this analysis and those given in B16 for targets common in both samples, as a function of $P_{\rm rot}$. Symbols are colour-coded by their $v\sin i$ measured values (given in GESiDR6) where black filled circles (i.e., corresponding to $v\sin i=0\,{\rm km\,s}^{-1}$) denote targets that are missing data.}
    \label{F_Bouvier_Comparison}
\end{figure}

Figure~\ref{F_Bouvier_Comparison}, which compares the distribution of $\delta$EW(Li) and $P_{\rm rot}$ between the B16 sample and our work, reveals two main points. Firstly, the B16 sample shows a weak, marginally significant negative correlation between $\delta$EW(Li) and $P_{\rm rot}$, in agreement with the findings of B16. However, the results from our independent analysis of stars in the same $T_{\rm eff}$ range with a larger sample is consistent with no correlation. If we return the 13 missing B16 targets into our sample, or only use the 50 targets in common with B16, but with our EW(Li), we find underlying weak but insignificant negative correlations.

The reason for any difference between the analysis in B16 and here does not appear to be a difference in the rotation periods used. There is only one of the 50 stars in common where the period has changed significantly. Instead the difference appears to be in the way that EW(Li) was determined from the same data (albeit the B16 analysis was based on the earlier GESiDR4 spectra). In the bottom panel of Figure~\ref{F_Bouvier_Comparison} we show a comparison of the EW(Li) values between this work and B16 as a function of rotation period. The differences are positive on average but there appears to be a correlation, although the significance is not strong ($p=0.12$). This correlation points to a difference in the way broadened spectra are considered in the two approaches. B16 performed simulations to check whether rotational broadening could result in the inclusion of blends with surrounding iron lines, which might lead to an overestimate of EW(Li) for the fast rotators and the correlations we see in Figure.~\ref{F_Bouvier_Comparison}. Their result was that overestimates were probably limited to 10-20m\AA. Nevertheless we would claim that our technique of extracting over a window tuned to the target $v\sin i$ after subtraction of an appropriately broadened template should account better for the depression of the surrounding continuum due to the presence of other broadened features.

If we fit the correlation in the bottom panel of Fig.~\ref{F_Bouvier_Comparison} with a straight line and ``correct'' the B16 EW(Li)s with this then any correlation of $\delta$EW(Li) with rotation disappears ($p=0.32$). Another interpretation is that {\it our} EW(Li) values have been systematically underestimated for the rapid rotators (or overestimated for the slow rotators). However, if we apply the correction in the opposite direction on our sample, we still do not find a significant correlation ($p=0.16$) correlation with rotation period for this sample.

Another possibility is that any weak correlation with rotation in our data has been obscured by differential reddening. However, the uncertainty injected into $\delta$EW(Li) by the $\sim 0.06$ scatter in inferred $E(B-V)$ (see Table~\ref{T_Clusters}) is not big enough to cause uncertainties larger than the EW(Li) errors and in any case is comparable to the $\pm 200$ K $T_{\rm eff}$ estimates used by B16.

\bibliographystyle{mnras}
\bibliography{bibliography}
\label{lastpage}
\end{document}